\documentclass[twocolumn]{aastex61}
\usepackage{natbib}
\pdfoutput=1
\newcommand{\ha}{H$\alpha$}

\newcommand{\lir}{$L\rm_{IR}$}

\newcommand{\smy}{$\rm M_\odot ~ yr^{-1}$}

\newcommand{\rtwo}{$R_{200}$}

\newcommand{\sized}{$ R_{24}/R_{d}$}

\newcommand{\lcs}{$Local \ Cluster \ Survey $}

\newcommand{\sers}{{\it S\'{e}rsic}}
\newcommand{\redshiftcut}{$0.0137 < z < 0.0433$}
\newcommand{\ntot}{224}
\newcommand{\nmemb}{105}
\newcommand{\nfield}{119}

\shorttitle{The Local Cluster Survey I}
\shortauthors{Finn et al.}

\begin{document}
\title{The Local Cluster Survey I: Evidence of Outside-In Quenching in Dense Environments}
 
\correspondingauthor{Rose A. Finn}
\email{rfinn@siena.edu}

\author{Rose A. Finn}
\affil{Department of Physics \& Astronomy, Siena College, 515 Loudon
  Road, Loudonville, NY,  12211, USA}

\author{Vandana Desai}
\affil{IPAC, Mail Code 100-22, Caltech, 1200 E. California
  Boulevard, Pasadena, CA, 91125, USA}

\author{Gregory Rudnick}
\affil{University of Kansas, Department of Physics and
  Astronomy, Malott Hall, Room 1082, 1251 Wescoe Hall Drive, Lawrence,
  KS, 66045, USA}

\author{Michael Balogh}
\affil{Department of Physics and Astronomy, University of
  Waterloo, 200 University Avenue West,
Waterloo, Ontario, N2L 3G1, Canada}
\author{Martha P. Haynes}
\affil{Cornell Center for Astrophysics and Planetary Science, 530 Space
   Sciences Building, 122 Sciences Drive, Ithaca, NY 14853, USA}

\author{Pascale  Jablonka}
\affil{Observatoire de l'Universit\'e de Gen\`eve, Laboratoire d'Astrophysique de l'Ecole Polytechnique F\'ed\'erale de
  Lausanne (EPFL), 1290 Sauverny, Switzerland}

\author{Rebecca A. Koopmann}
\affil{Department of Physics \& Astronomy, Union College,
  Schenectady, NY, 12308, USA}

\author{John Moustakas}
\affil{Department of Physics \& Astronomy, Siena College, 515 Loudon
  Road., Loudonville, NY, 12211, USA}

\author{Chien Y. Peng}
\affil{Giant Magellan Telescope Observatory, 465 N. Halstead
  Street, Suite 250, Pasadena, CA, 91107, USA}

\author{Bianca Poggianti}
\affil{Osservatorio Astronomico, vicolo dell'Osservatorio 5, 35122
  Padova, Italy}

\author{Kenneth Rines}
\affil{Department of Physics \& Astronomy, Western
  Washington University, Bellingham, WA 98225, USA}

\author{Dennis Zaritsky}
\affil{Steward Observatory, 933 N. Cherry Avenue, University of Arizona,
  Tucson, AZ, 85721, USA}

\begin{abstract}
{
The goal of the {\it Local Cluster Survey} is to look for evidence of
  environmentally driven quenching among star-forming galaxies in 
  nearby galaxy groups and clusters.  Quenching is linked with
  environment and stellar mass, and much of the current observational
  evidence comes from the integrated properties of galaxies. However,
  the relative size of the stellar and star-forming disk is sensitive
  to environmental processing and can help identify the mechanisms
  that lead to a large fraction of quenched galaxies in dense
  environments. Toward this end, we measure the size of the
  star-forming disks for 224 galaxies in nine groups and clusters ($0.02<z<0.04$;
$\rm SFR > 0.1$~\smy) using 24\micron \ imaging from the {\it Spitzer  Space
 Telescope}.  We normalize
the 24\micron \ effective radius ($R_{24}$) by the size of the stellar
disk ($R_d$).   We find that star-forming galaxies with higher
bulge-to-total ratios ($B/T$) and galaxies in more dense environments
have more centrally concentrated star formation. 
Comparison with H~I mass fractions and ${\rm NUV}-r$ colors
indicates that a galaxy's transition from gas-rich and blue to depleted
and red is accompanied by an increase in the central concentration of
star formation.  We build a simple model to constrain the timescale over which the
star-forming disks shrink in the cluster environment. Our results are
consistent with a long-timescale ($>2$~Gyr) mechanism that produces
outside-in quenching, such as the removal of the extended gas halo or weak stripping of the cold disk gas.
}

\end{abstract}


\section{Introduction}
\label{introduction}

Galaxy clusters host a smaller fraction of
actively star-forming galaxies than the general field
\citep[e.g.][]{balogh97,poggianti99,lewis02,gomez03,finn05,finn08, finn10,postman05}.  
Even at fixed stellar mass, the fraction of star-forming galaxies tends to
decrease with increasing environmental density \citep[e.g.][]{peng10}.
The mass accretion rate of clusters
\citep[e.g.][]{mcgee09}, combined with the assumption that all accreted galaxies
eventually quench their star formation, implies that 
quenching must take place over a long timescale, 3--7 Gyr 
\citep[e.g.][]{balogh00, ellingson01, kimm09, peng10, delucia12, wetzel12}.

However, trying to uncover the physics that drives this transformation
by studying the properties of galaxies during the transition has
proved difficult and controversial.  A small fraction of galaxies are
found in a post-starburst phase, which implies a rapid truncation
timescale \citep{dressler83}; this fraction may depend on environment
and redshift 
\citep[e.g.][]{zabludoff96,balogh99,poggianti99,tran07,mok13,muzzin13}. 
However, the bulk of the star-forming population in clusters looks very
similar to those in the field, in terms of their star formation rate
and color distribution 
\citep{balogh04, baldry06, mcgee11}, with just the relative fraction of
star-forming and quiescent galaxies varying with environment.  
Few galaxies, at least at low
redshift, seem to inhabit the green valley region that separates
star-forming from quiescent galaxies, which implies that the
transition itself must happen rapidly, on timescales $<1$~Gyr
\citep{wetzel12}.  

To reconcile these facts, \citet{wetzel12, wetzel13} 
proposed a two-stage model in which star formation truncation happens
rapidly, but only after a lengthy delay time following accretion into
a cluster or group.  In the model as originally proposed, 
galaxies experience little or no change in their star-formation rates
during the delay phase.
Understanding how galaxies could remain uninfluenced by their
environment for so long has been a theoretical challenge.
Nonetheless, the model has been used successfully to interpret a range of
observations at redshifts out to $z=1$
\citep{vanderburg13, mok14, muzzin14, haines15, knobel15, fossati17}.
Following the delay, the process of galaxy quenching proceeds on a
rapid timescale of $<1$~Gyr.

Despite the success of this delay+rapid quenching model, the
hypothesis that galaxies are completely unaffected during the delay
period remains controversial.  Several studies 
have shown evidence that at a fixed stellar mass, star-formation rates of galaxies in dense
environments are skewed to lower values than in the general field 
\citep[e.g.][]{koopmann04a,poggianti08, gallazzi09, wolf09, finn10,vulcani10, haines13, lin14, taranu14, rodriguez-del-pino17}.  
This would imply that there {\it are} environmental processes at work
during the several Gyrs after infall.  The delay period might then
simply be a time during which long-timescale processes are slowly
altering the cluster galaxies, prior to the ultimate termination of
star formation in a rapid quenching phase \citep[e.g.][]{haines15}.

In this paper, we investigate what is happening to galaxies
during the delay phase, and we identify delay-phase galaxies as
those that have been accreted by a cluster but are still forming
stars.  Our goal is to measure the relative extent of the gas and
stellar disks, because this is a sensitive probe of environmental
processing \citep[e.g.][]{moss00, dale01, koopmann06b, bamford07,kawata08, jaffe11, bosch13, bekki14, schaefer17}. 
We present a sample of \ntot \ galaxies in 9 low-redshift
galaxy groups and clusters, 
and we map galaxies from the cluster core
to the surrounding infall regions.  
Our galaxy sample spans
a large range in stellar mass, which
is necessary to help control for any intrinsic quenching mechanisms
that may occur independently of environment \citep[e.g.][]{peng10}.  
We
use 24\micron \ imaging from the MIPS camera on $Spitzer \ Space \
Telescope$ to measure the size of the star-forming disks, and  
SDSS $r$-band imaging to quantify the size of the stellar disk.

Multiple wavelengths can be used to trace the star-forming regions in
galaxies, including UV, \ha, and infrared emission. While
24\micron \ is a reliable star-formation indicator that closely traces
Paschen alpha \citep[e.g.][]{calzetti07}, different wavelength
indicators can vary with dust and metallicity,  and may
therefore result in different measurements of the spatial extent of
the gas disk.  Furthermore, different components of the ISM will
respond differently to environmental processing
\citep[e.g.][]{boselli14}.  To mitigate systematics associated with
our choice of wavelength, we normalize the size of the star-forming
disk by the size of the stellar disk. Our conclusions are based upon
{\em differences} in this ratio with environment and intrinsic galaxy properties, rather than absolute measurements.

This paper is organized as follows.
We describe the survey and cluster properties in Section \ref{lcs}.
In Section \ref{obs} we describe the $Spitzer$ observations, and in
Section \ref{selection} we detail the selection and
properties of the galaxies in our sample.  In Section \ref{env}, we
present our methods for quantifying
the environment of galaxies, and we discuss our measurements of 24\micron
\ sizes in Section \ref{image}.  Finally, we present our results in
Section \ref{results}, which show that the properties of star-forming galaxies are
altered during the delay, prior to their final quenching.   
When calculating distance-dependent quantities, we use $H_0=70 \rm~km~s^{-1}$, $\Omega_\Lambda=0.7$, and 
$\Omega_M=0.3$.  Stellar masses are calculated as described in 
\citet{moustakas13} and assume a Chabrier
initial mass function \citep{chabrier03}.

\section{The Local Cluster Survey} 
\label{lcs}
\begin{figure}
\includegraphics[width=.5\textwidth]{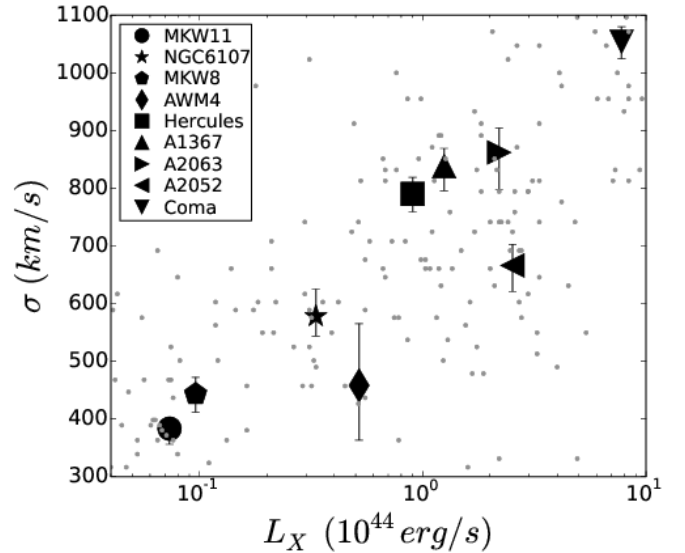}
\caption{Cluster biweight scale versus X-ray luminosity for
  Local Cluster Survey sample (large points).  The cluster sample purposely spans a
  large range in X-ray luminosity so that we sample the full range of
  galaxy environments.  The small points show a larger sample of
  clusters from \citet{mahdavi01} to illustrate that the scatter seen among \lcs \ clusters is consistent with 
the scatter seen among a larger population of clusters. }
\label{sigmalx}
\end{figure}

The nine clusters that compose the \lcs \ are listed in Table \ref{finalsample}.
The sample consists of clusters that have wide-area $Spitzer$  MIPS
24\micron \ mapping and that 
lie in the SDSS \citep{york00} and ALFALFA \citep{giovanelli05} 
surveys.  The clusters are near enough
so that a typical spiral can be resolved in a $Spitzer$ 24\micron 
\ image yet far enough so that the SDSS photometry is reliable (\redshiftcut).
The clusters purposely span a range of richness, X-ray luminosity, and
X-ray temperature so that we can probe the full range of intra-cluster 
medium properties.  In Figure \ref{sigmalx}, we show X-ray luminosity
{in the $0.1 - 2.4$~keV band} versus cluster velocity 
dispersion for our sample.  { For comparison, we show a larger
  sample of clusters from \citet{mahdavi01} with gray points to
  illustrate that the scatter seen among \lcs \ clusters is consistent with 
the scatter seen among a larger population of clusters. }

While Figure \ref{sigmalx} demonstrates the range in global cluster environments, the
galaxies in our sample also span a wide range of local
environments.  
The galaxies in the MIPS scans of the lower-mass clusters sample both
low and high density regions,
whereas the galaxies in the MIPS scans of Coma, Hercules, and Abell 1367 provide a more complete sampling of
the highest-density environments.  An important point to keep in mind as we proceed with the analysis is
that Coma galaxies dominate the number counts at high local densities.  
We investigate how this impacts our results in Section \ref{comaresults}.

\begin{figure*}
\includegraphics[width=.7\textwidth]{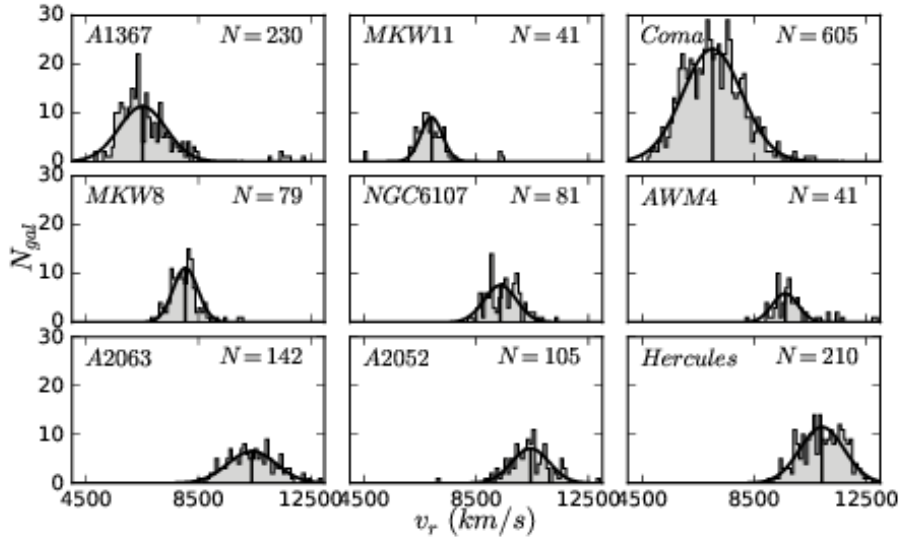}
\caption{Velocity distribution of NSA galaxies used for the
  calculation of biweight location and scale.  The clusters are
  ordered by central recession velocity.  In
  each panel, the resulting number of cluster members is indicated in
  the top right.  The
  vertical black line shows the biweight location, and the black curve
  shows a gaussian distribution centered on the biweight location,
  with a width equal to the biweight scale.  $N$ is the number of
galaxies with velocity offsets $\Delta v < 3
\sigma$ and that have projected radii less that
\rtwo.
\label{biweight_hists} }
\end{figure*}

While some clusters have average redshift and velocity dispersions available from the 
literature, we recalculate the biweight location and scale \citep{beers90} and
estimate errors using bootstrap resampling.  
The galaxies for these calculations and throughout 
are drawn from the NASA-Sloan Atlas \citep{blanton11}, 
which includes
SDSS spectroscopic sources along with other spectroscopically confirmed
galaxies from ALFALFA and other surveys.  We select all galaxies that
have redshifts between \redshiftcut, where the lower limit is set
by Coma (recession velocity minus 3$\sigma_v$) and the upper limit is
set by the most distant cluster, Hercules (recession velocity plus 3$\sigma$).  
We first calculate the biweight location and scale 
using all galaxies, with a recession velocity within 4000~$\rm km~s^{-1}$ of each cluster and a 
projected radius less than
1.7~Mpc (this corresponds to a 1 degree radius at the redshift of
Coma).  
We use this biweight location as the new median and
recalculate the location and scale.  We repeat this until the scale
changes by less than 1~$\rm km~s^{-1}$, which usually happens within two
iterations.  We use the weighting factors suggested by
\citet{beers90} to minimize the effect of any galaxy whose
velocity deviates by more than 4$\sigma_v$ from the central velocity.  
We show the results in Figure \ref{biweight_hists} and 
in Table \ref{finalsample}.  The errors listed in Table
\ref{finalsample} are the 68\%
confidence interval as determined by bootstrap resampling.
In each panel in Figure \ref{biweight_hists}, we list the number of
galaxies with velocity offsets $\Delta v < 3
\sigma$ and that have projected radii less that
\rtwo, where \rtwo \ is calculated from the cluster biweight scale
according to the relation in \citet{finn08}.  

\begin{deluxetable*}{ccccc} 
\tablecaption{Cluster Properties and Galaxy Sample Sizes  \label{finalsample}} 
\tablehead{\colhead{Cluster} &\colhead{Biweight Central Velocity}  &
  \colhead{Biweight Scale} & \colhead{$N_{gal}$} & \colhead{$N_{gal}$}
  \\ & \colhead{($\rm km~s^{-1}$)}  & \colhead{($\rm km~s^{-1}$)}  & Core & External } 
\startdata 
A1367 & 6505$^{+55}_{-54}$  & 838$^{+31}_{-42}$  & 6 & 5  \\ 
MKW11 & 6904$^{+38}_{-49}$  & 383$^{+19}_{-27}$  & 5 & 13  \\ 
Coma & 7011$^{+45}_{-44}$  & 1054$^{+26}_{-29}$  & 28 & 11  \\ 
MKW8 & 8039$^{+40}_{-38}$  & 443$^{+29}_{-31}$  & 0 & 11  \\ 
NGC6107 & 9397$^{+57}_{-53}$  & 578$^{+47}_{-34}$  & 6 & 17  \\ 
AWM4 & 9604$^{+61}_{-55}$  & 458$^{+107}_{-95}$  & 4 & 17  \\ 
A2063 & 10410$^{+72}_{-74}$  & 862$^{+42}_{-65}$  & 27 & 8  \\ 
A2052 & 10431$^{+57}_{-64}$  & 666$^{+37}_{-45}$  & 7 & 34  \\ 
Hercules & 10917$^{+50}_{-53}$  & 790$^{+29}_{-31}$  & 22 & 3  \\ 
\enddata 
\end{deluxetable*}

\section{MIPS Observations and Reduction}
\label{obs}
\begin{figure*}[h]
\includegraphics[width=\textwidth]{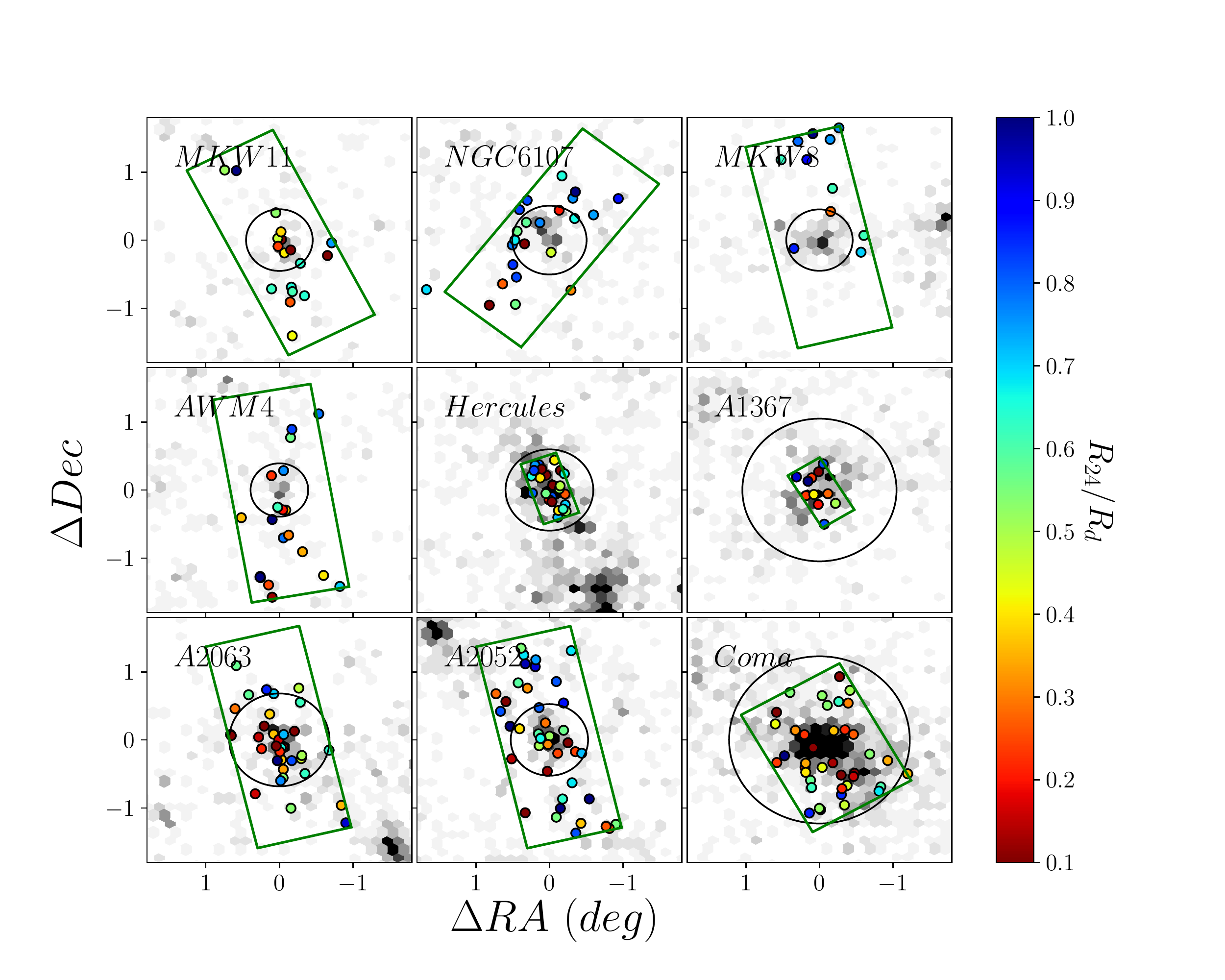}
\caption{Dec versus RA for galaxies in the vicinity of each cluster.  
Clusters are ordered left-to-right and top-to-bottom by increasing X-ray
luminosity.  The colored circles show the projected position of
galaxies in the GALFIT sample, with 
the color indicating \sized.
The grayscale shows the surface density of galaxies
within $\pm3 \sigma_v$, and the large black circles show \rtwo.  
The green rectangle shows the footprint of the 24\micron \ MIPS scan.
\label{plotpositions24}}
\end{figure*}

Each cluster has 24$\mu$m data from the MIPS instrument 
\citep{rieke04} on the {\it Spitzer  Space  Telescope}, 
and we use the 24\micron \ emission to probe the spatial extent of
star formation.
In Figure \ref{plotpositions24}, we show the positions of galaxies
that lie within the 24\micron \ scan region and have a 
redshift in the range of $0.01366 < z < 0.04333$.
The overall galaxy density is shown with the grayscale, with black and white 
denoting high and low density areas, respectively.  Specifically, white
indicates bins with no galaxies, and black indicates regions that
contain 10 or more galaxies.
The dark gray circles show $R_{200}$, and the green box shows the
footprint of the MIPS scan.
With the exception of Coma, A1367, and Hercules, the clusters have
MIPS data obtained specifically for this project,
and each scan covers an area of approximately
$1.5 \times 2.5$ square-degree area around each cluster.  
The MIPS data for 
Coma, A1367, and Hercules were pulled from the {\it Spitzer Science
  Center} archive, and the observations are summarized in Table \ref{observations}.
While the areal coverage for these
clusters is smaller, the archive clusters provide an important 
complement by sampling regions of higher local density and
X-ray luminosity (see Figure \ref{sigmalx}).

\begin{deluxetable*}{lccccccc}
\tablecaption{$Spitzer$ MIPS Observations \label{observations}}
\tablehead{\colhead{Cluster} & \colhead{$z$} & \colhead{Date of Obs}
  & Program ID & PI & \colhead{$f_{80} \rm (MJy/sr)$} & \colhead{$\log_{10}(L_{80}/L_\odot)$}  }
\startdata
A1367    & 0.0217 & 2006 Jun 07 & 25 & Fazio & 3.50$\pm$0.25 & 8.25\\
MKW11    & 0.0230 & 2008 Jul 31 & 50456 & Finn & 4.00$\pm$0.25 & 8.38\\ 
Coma     & 0.0234 & 2004 Jun 22 & 83 &Rieke, G. & 2.25$\pm$0.25 & 8.12\\
MKW8     & 0.0268 &2009 Mar 24 & 50456 & Finn & 3.75$\pm$0.25 & 8.45\\ 
NGC6107  & 0.0313 &2008 May 16 & 50456 & Finn & 3.25$\pm$0.25& 8.50\\ 
AWM4     & 0.0320   &2008 Sep 26 & 50456 & Finn & 3.50$\pm$0.25 & 8.56\\ 
    & &   2008 Sep 28 \\ 
A2063    & 0.0347 &2009 Mar 23 & 50456 & Finn & 4.00$\pm$0.25 & 8.71\\ 
A2052    & 0.0348 &2009 Mar 24 & 50456 & Finn & 4.00$\pm$0.25 & 8.71\\ 
Hercules & 0.0364 & 2006 Jun 22 & 25 & Fazio &3.25$\pm$0.25 & 8.68\\
\enddata
\end{deluxetable*}

The MIPS data are reduced using the MOPEX software package, following
the procedure outlined in the $Spitzer$ MIPS Data Reduction Cookbook\footnote{http://irsa.ipac.caltech.edu/data/SPITZER/docs/dataanalysistools/cookbook/home/}.
We use SExtractor \citep{bertin96} to detect galaxies and measure photometry.  We
measure the noise for each object using the MOPEX-generated
standard deviation image.
We estimate the depth of each image by placing artificial galaxies on each image
and then rerunning SExtractor.  We calculate the flux where we recover 80\% of the
test sources, and we list these flux limits in Table \ref{observations}.  
We convert the 80\% flux limits to an IR luminosity at the cluster redshift using the
templates of \citet{chary01} and list these values in the final column
of Table \ref{observations}.

\section{Selection of Star-Forming Galaxies}
\label{selection}

Our parent sample consists of exactly 1800 NSA galaxies that lie on the MIPS
24\micron \ scans and fall in the redshift range \redshiftcut. 
We apply additional selection criteria to ensure that we are sampling the
same galaxy population in each cluster.
First, to account for the varying sensitivity of the
24\micron \  scans, we apply a uniform cut in $L_{IR}$ ($L_{\rm IR} > 5.2 \times
10^8~L_\odot$).  This limit is set by the depth of the MIPS scans for 
Abell~2052 and Abell~2063, which have the highest 80\%
completeness limit of $L_{\rm IR} = 5.13 \times 10^8~L_\odot$.  
Assuming that the 24\micron \
emission is due to heating from massive stars, this \lir \  corresponds to an 
approximate star-formation rate of 0.1~\smy.    
We retain 541 
galaxies after the $L_{\rm IR}$ cut.

We use both optical emission-line ratios
and infrared colors to eliminate AGN from our sample.  The majority of
galaxies in the NASA-Sloan Atlas have
optical spectra, 
and we use the \citet{kauffmann03} 
criteria for separating AGN from star-forming galaxies.
For those without optical spectra, we use $WISE$ photometry 
$W1-W2 > 0.8$ to identify AGN \citep{stern12}. (Note
that only 10/1800 galaxies in our sample show such red $W1-W2$
colors.  This is likely because this color selects only whopping AGN.)
Two galaxies have neither optical spectra nor $WISE$ photometry, and so 
we are not able to classify as AGN vs. star
forming; we retain them in the sample nonetheless.  After removing
AGN, we are left with 351 star-forming galaxies.

 We
also apply a minimum cut on galaxy size and keep only galaxies with
$r$-band effective radii $R_e \ge 1.3$~kpc.  This corresponds to 1
pixel on the MIPS detector at the distance of our farthest cluster,
Hercules ($2\arcsec .45 = 1.3$~kpc), and we are not confident in
measuring sizes that are smaller than a pixel.  
We retain  332 of the galaxies
after applying the size selection.  

We attempt to model all of these galaxies using GALFIT
\citep{peng02}.  
We remove six galaxies (three pairs) that are
close enough to each other to compromise the resulting fits.  
The modeling fails for another 40 galaxies, and this is usually due to the
low signal-to-noise ratio of the 24\micron \ emission.  
In addition,
we remove 33 galaxies with observed 24\micron \ surface brightnesses greater than 20
$\rm mag/arcsec^2$ because our simulations indicate that the fits are
unreliable; we discuss the details of these simulations in the Appendix
\ref{sim}.
We retain 252 galaxies.

Finally, we require galaxies to be in the sample of
\citet{simard11} because we use their measurements of effective radius
and bulge-to-total ratio in our analysis.
\citet{simard11} perform two-dimensional
bulge-to-disk decomposition for more than 1 million galaxies in
SDSS DR7 using the GIM2D software.  They limit their sample to
galaxies with $14 \le m_{\rm petro,r,corr} \le 18$ and exclude objects
that are classified in the DR7 photometry table as saturated or
unresolved.  We are able to match $89\%$
of our sources to the \citet{simard11} catalog, which leaves us with
a final sample of \ntot \ galaxies.  Of the 28 galaxies not matched to
the \citet{simard11} catalog, 17 are too bright ($m_r < 14$), two are too
faint ($m_r > 18$), two are blended, two are saturated, one is not in
the DR7 catalog, and one seems to have bad coordinates in the DR7
catalog.  We are not able to identify the reason why
the remaining three galaxies are not in the Simard catalog.

\section{Quantifying the Environment of Star-Forming Galaxies}
\label{env}

\begin{figure}[h]
\includegraphics[width=.48\textwidth]{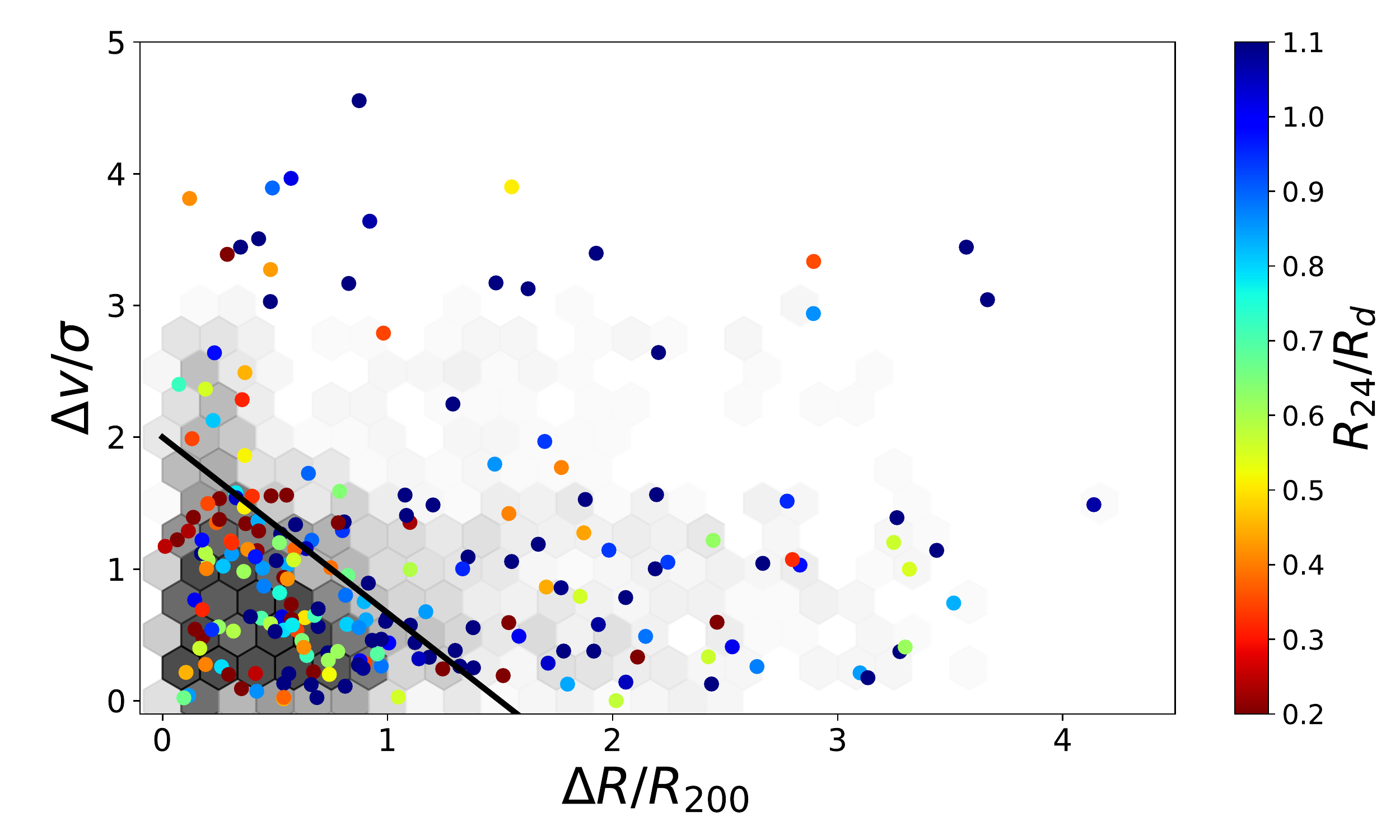}
\caption{$\Delta v/\sigma$ versus $\Delta R/R_{200}$ for galaxies in the
  MIPS scans.  The colored points show the GALFIT galaxy sample, and the
  color indicates the relative size of the 24\micron \ and $r$-band
  half-light radii.  The grayscale shows the phase-space distribution
  of both core and external galaxies that lie within the MIPS areal
  coverage.  The black line shows how we separate core and external
  galaxies \citep{oman13}.  Seven additional galaxies lie at $\Delta
  v/\sigma > 5$ but are not shown here, so that galaxies with lower
  velocity offsets can be seen more clearly.  
Note that the external sample likely contains galaxies
  that have already passed through the cluster center.  }
\label{phasespace}
\end{figure}

{We will compare the size of the star-forming disk versus environment
throughout this paper, and we define environment in two different but
related ways.  First, we make a simple division in phase space to
split the sample into two groups, galaxies in the cluster core
versus galaxies external to the core.
We adopt the cut used by \citet{oman13},
$\left |{\Delta v/\sigma}\right | < -4/3 \Delta
R/R_{200} + 2$, and we show this cut with the black line in Figure \ref{phasespace}.
This figure includes galaxies from all clusters, and the black shading shows the
phase-space positions of all galaxies.  The colored points show the
GALFIT sample  (\S \ref{image}).  
The region in phase space to the left of the black line contains galaxies
that are likely to be true cluster members, whereas the region to the
right of this line contains galaxies near the cluster as well as a
large fraction ($>50$\%) of interlopers that are not physically associated with
the cluster \citep{oman16}.
We denote the sample to the left and right of this line as core and
external galaxies, respectively.
According to this
definition, our sample contains \nmemb \  core galaxies and \nfield \ 
external galaxies, and the breakdown of galaxies by cluster is shown in columns 4 and 5 of
Table \ref{finalsample}.

We note that neither the core nor external samples are pure.  The core
sample is dominated by galaxies that have been in the cluster
environment the longest \citep[e.g.][]{oman13} but may still contain galaxies that are physically distant from the cluster
center but lie along the line of sight.  In addition, the external
sample will contain galaxies in the field as well as backsplash
galaxies that have already passed through the cluster core  \citep[e.g.][]{balogh00,bahe13}.
Therefore core/external do not translate cleanly into cluster/field,
and this cross-contamination will dilute the observational
signatures of any environmental processing.

}

\begin{figure}[h]
\includegraphics[width=.5\textwidth]{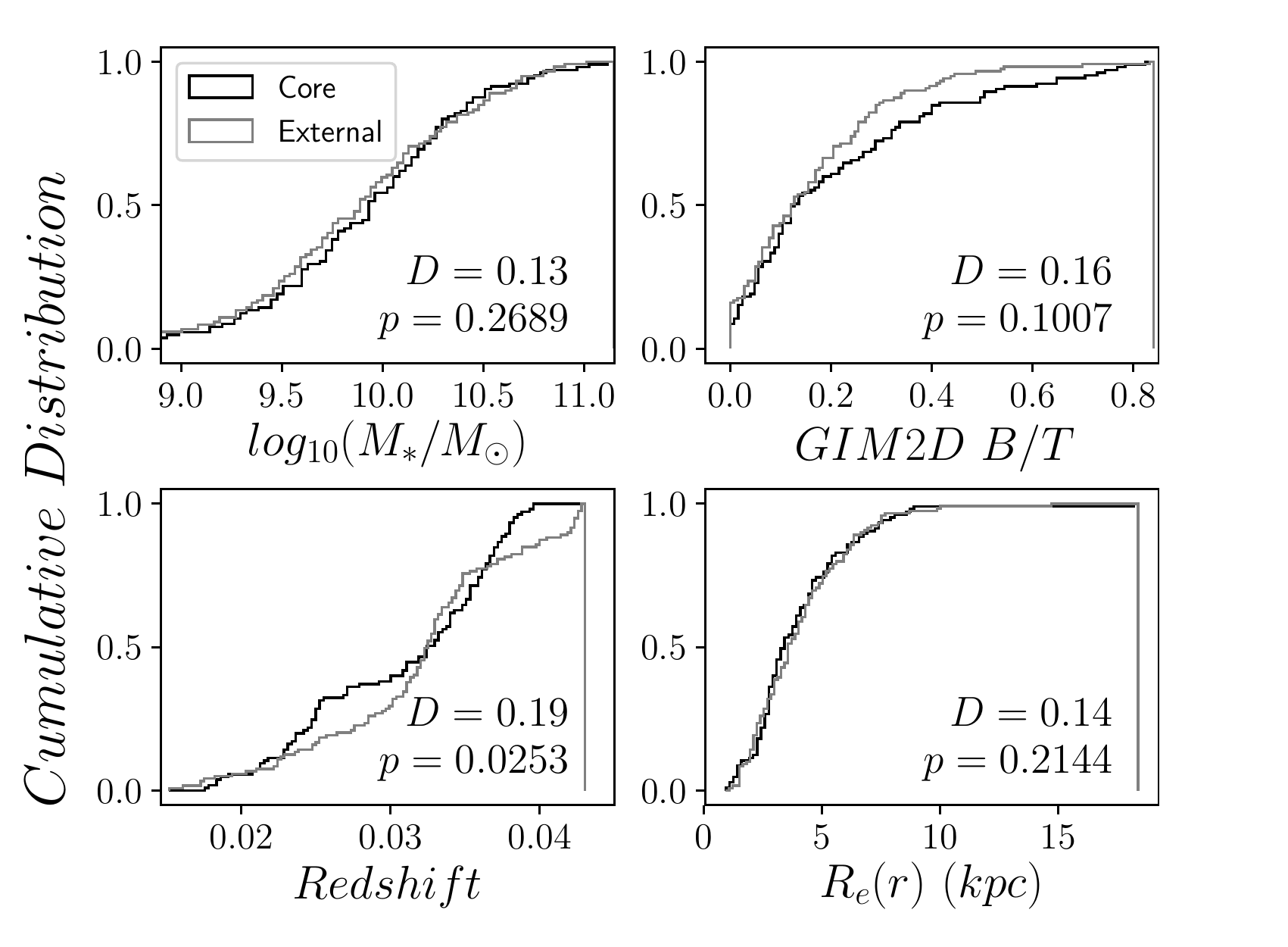}
\caption{Cumulative distributions of stellar mass, $B/T$, redshift, and effective
  radius for core (black) and external galaxies (gray).  The K-S D and $p$-value are
  shown in each panel.  The 
  samples are comparable except in terms of redshift, with the external sample
  offset to slightly higher values.  This difference does not affect
  our results.}
\label{clusterfield}
\end{figure}

To demonstrate the similarity and therefore comparability of the
core and external subsamples, we show the
cumulative distributions for stellar mass, bulge-to-total ratio, 
redshift, and $r$-band effective radius in Figure
\ref{clusterfield}.  The properties of the core and external galaxies
are not statistically different according to both a K-S and
Anderson-Darling test, except with regard to redshift.  {Coma
contributes significantly to the core sample, whereas the redshift
distribution of the external sample is skewed
slightly toward higher redshift.}  This difference does not impact our results.

{We use local galaxy density as a second metric of galaxy environment.
Specifically, we use $\Sigma_5$, which is the surface density of galaxies out to the fifth nearest neighbor.
When counting neighbors, we use only galaxies in the SDSS spectroscopic sample with $M_r
< -18.2$.  The magnitude cut corresponds to the SDSS spectroscopic completeness
limit ($ r < 17.7$) at the distance of our farthest cluster.  In
addition, we require that neighbors must have a velocity offset
$\Delta v < 2500~\rm km~s^{-1}$.

We note that our two measures of environment are closely related; $\Sigma_5$ is strongly correlated with a galaxy's position in phase
space.  In effect, $\Sigma_5$ provides a continuous measure of
environmental density as compared to the dichotomous division
of core versus external galaxies.  

Two clusters in our sample, A2052 and A2063, lie near each other on
the plane of the sky and in redshift space, but we are still confident
in our ability to quantify the environment of galaxies in this
region.  The core/external division is preserved around
these two structures because while the MIPS scans overlap slightly, there are no galaxies that are classified
as external in one field of view and core in another field of view, or
vice versa.  In
addition, the local density metric will account for any 
enhanced density between the two clusters (see Figure \ref{plotpositions24}).
}

\section{Analysis of Optical and IR Images}
\label{image}
Our goal is to measure the size of each galaxy in the optical and
infrared to trace the stellar and star-forming components, respectively.
\citet{simard11} 
use GIM2D to fit two-dimensional, single-component \sers \ model for a
large fraction of DR7 galaxies, and they also fit two versions of
bulge$+$disk models to each galaxy.  One version of the bulge$+$disk models forces
the bulge to have an $n=4$ \sers \ profile, and the second set of
models allows the \sers\ index of the bulge to vary.  
We use the $r$-band disk radius ($R_d$; see Table \ref{radii} for
a list of definitions of radial size measurements) for the $n=4$ bulge$+$disk models to characterize the size
of the stellar disk.
We use GALFIT \citep{peng02} to fit a \sers \ model to the 24\micron \ images.
To confirm that GALFIT and the GIM2D models are comparable, we 
fit the $r$-band images for a subset of the galaxies using 
GALFIT and find that the GIM2D and GALFIT model parameters are consistent.

\subsection{GALFIT Analysis of 24\micron \ Images}
\label{imagefitting}
To quantify the 24\micron \  size of each galaxy, 
we fit a two-dimensional, single-component \sers \ profile to the
24\micron \ galaxy images using GALFIT \citep{peng02}.  
GALFIT requires a PSF image to properly model galaxies, and this is
particularly important when modeling low-resolution data such as the
24\micron \ scans.  We use a point source in each
MIPS mosaic as the reference PRF for that field, 
choosing a bright, isolated source that shows a
clear diffraction pattern.  For Hercules, we use the PRF provided by
$Spitzer \ Science \ Center$ because our simulations indicate that the
PRFs we created were biasing the model parameters (see the Appendix \ref{sim}).
We make a $30\times30~\rm pixel^2$ cutout, keeping the star centroid at
$(16,16)$ as required by GALFIT.  To test the accuracy of the PSF, we 
select other stars on the image and perform a one-component fit using GALFIT.  
We are able to model the other point sources well, with 
little residual remaining after subtracting the model from the image.
The model parameters are $R_e \approx 2 ~\rm pixels$, \sers \ index $\approx
1.5$, and $B/A=0.9-1.0$.  
The fitted 
magnitudes are comparable to those measured by SExtractor MAG\_BEST.

In addition to the PRF image, GALFIT requires an input image, a
noise image, and a mask if multiple objects fall within the analysis area.
We create cutouts for each galaxy from the MIPS scan, using an analysis area of 100\arcsec$\times$100\arcsec \ or
six times the $r$-band Petrosian $R_{90}$ (whichever is larger).  
We mask out other objects in the analysis region using a segmentation
image from SExtractor, as suggested in the GALFIT {\it User
  Manual}\footnote{ http://users.obs.carnegiescience.edu/peng/work/GALFIT/GALFIT.html}.
We review these masks by hand and fix the masks for galaxies that were broken 
into several objects by SExtractor; this is typically an issue for
larger, well-resolved spirals with more clumpy features.
We use the standard deviation image that is created by MOPEX  as 
the noise image for GALFIT.   

We run GALFIT on each 24\micron \ image using a one-component \sers \
model.  We use the GIM2D one-component $r$-band \sers \ model parameters as an initial guess for 
the 24\micron \ \sers \ fit.  Because most galaxies are only marginally
resolved at 24\micron, we hold the axis ratio and
position angle fixed at the $r$-band values.  If the best-fit \sers \ index
goes above 6,  we refit the galaxy while
holding the \sers \ index fixed at 5.  
{In some cases, the GALFIT model converges with a numerical
error.}  In most of these cases, the error arises because the 24\micron \
emission is small and consistent with a point source, and the effective radius of
the model galaxy is close to zero.  For some of the fainter point-like sources, a model will
converge without a numerical error if we let the axis ratio vary.  If we are unable to obtain a
reliable fit,  we remove these galaxies from the sample.

\subsection{GALFIT Best-fit Model Parameters}

\begin{figure}[h]
\includegraphics[width=.5\textwidth]{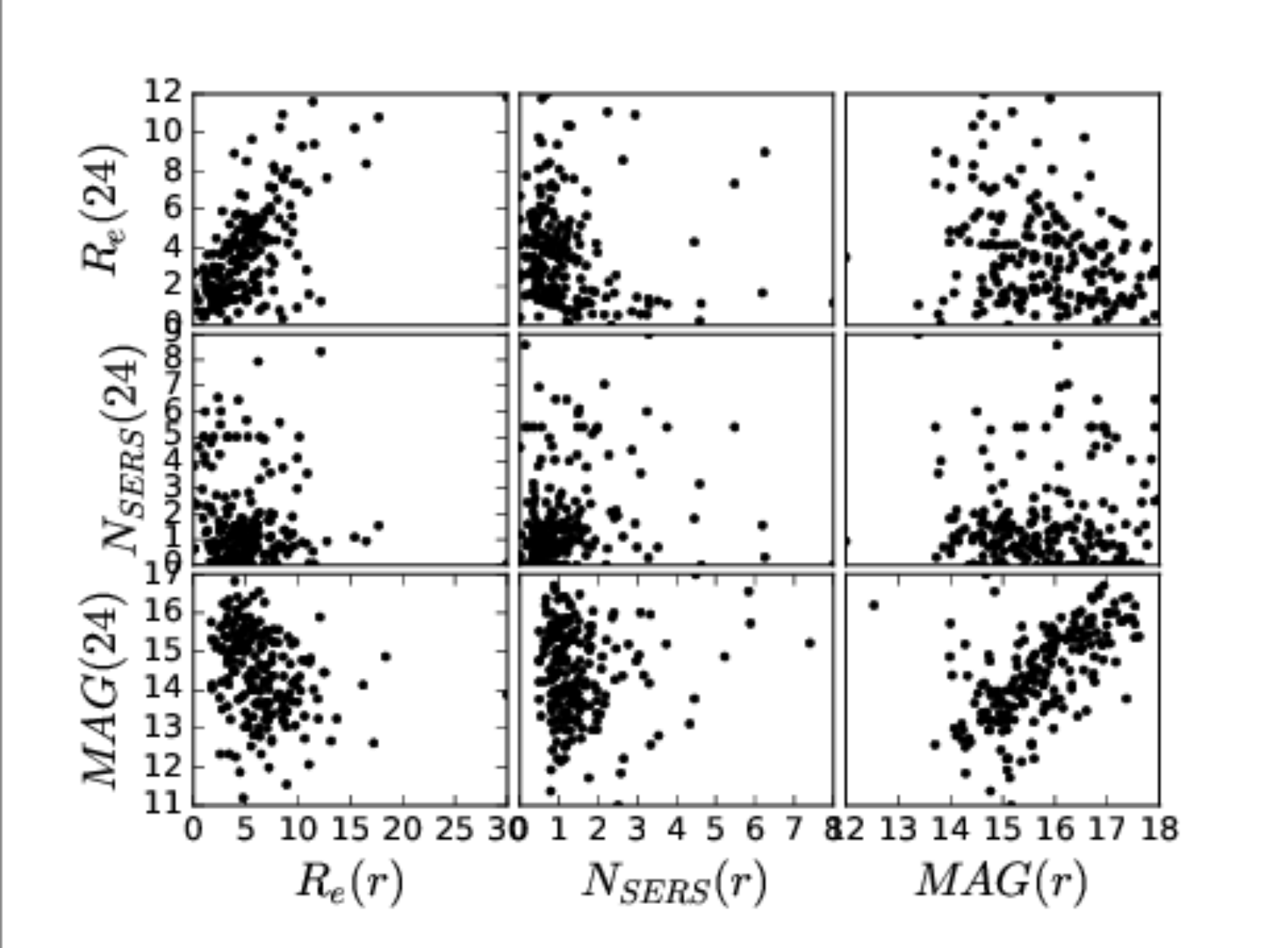}
\caption{\small   GALFIT 24\micron \  model parameters versus \citet{simard11} $r$-band 
  model parameters for all galaxies in our final sample.  The top-left
and bottom-right panels show that 24\micron \ and $r$-band effective radius and
magnitude are strongly correlated.}
\label{GALFITvsNSA}
\end{figure}

We compare the 24\micron \ 
and the $r$-band \sers \ model parameters in Figure \ref{GALFITvsNSA}.
The $r$-band parameters are shown along
the horizontal axes, and the GALFIT 24\micron \ parameters are shown
along the vertical axes. The diagonal panels show
the corresponding parameters of the $r$ and 24\micron \ fits. The $r$
and 24\micron \ effective radii are
correlated, as are the model magnitudes.  The 24 and $r$ \sers \
indices are not strongly correlated, although the bulk of both $r$ and 24 indices
are less than 2.  This is consistent with a sample dominated by disk galaxies.  The top
middle panel shows that galaxies with higher $r$-band \sers \ index
have systematically lower values of $R_{24}$, a point we will revisit
in Section \ref{sizestuff}.  We conclude that there are no strange 
systematics between the 24\micron \ and $r$-band model parameters.
We discuss the reliability of the GALFIT model parameters in more
detail in the Appendix.

\section{Results}
\label{results}

\subsection{Relative Sizes of the Star-Forming and Stellar Disks}
\label{relativesize}

\begin{figure*}[h]
\includegraphics[width=\textwidth]{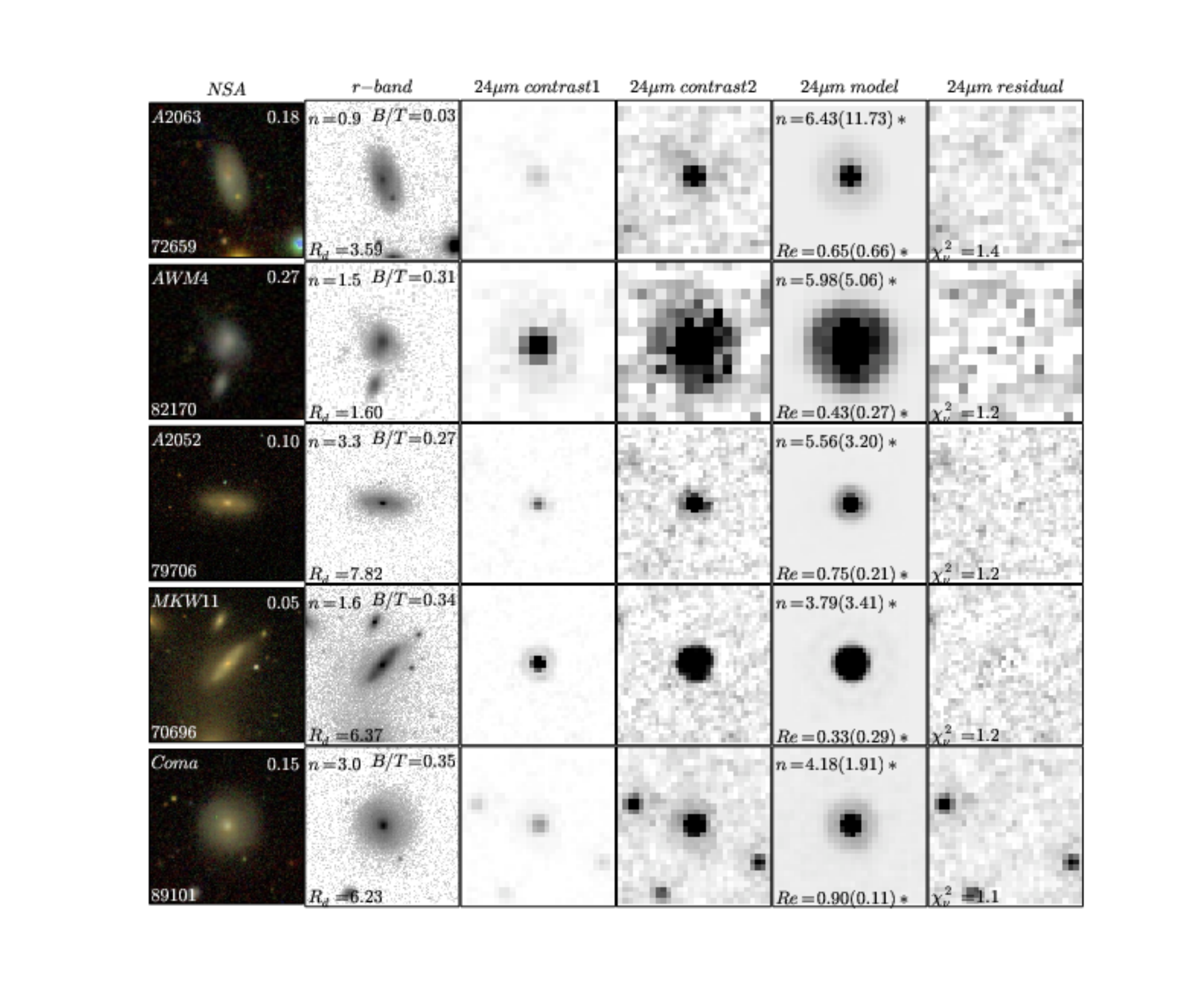}
\caption{\small Example GALFIT modeling results for
  five randomly selected galaxies with \sized \ between 0 and 0.3.  Columns 1 and 2 show the
  SDSS color and $r$-band images.  Columns 3 and 4 show the 24\micron
  \ image at two different stretches to emphasize the high and low
  surface-brightness features, respectively.  Columns 5 and 6 show the
  24\micron \ single-component \sers \ model
  and residual, and the contrast is identical to column 4 to emphasize the low
  surface-brightness features.
The text in column 1 gives the parent cluster, the NSA ID, and
\sized. Column 2 lists the $r$-band \sers \ index n, the
bulge-to-total ratio, and the disk scale length in arcseconds.  The text in column 5
lists the 24\micron \ \sers \ index and effective radius in arcseconds, and column 6
lists $\chi^2_\nu$ of the best-fit model.}
\label{smallsize}
\end{figure*}

\begin{figure*}[h]
\includegraphics[width=\textwidth]{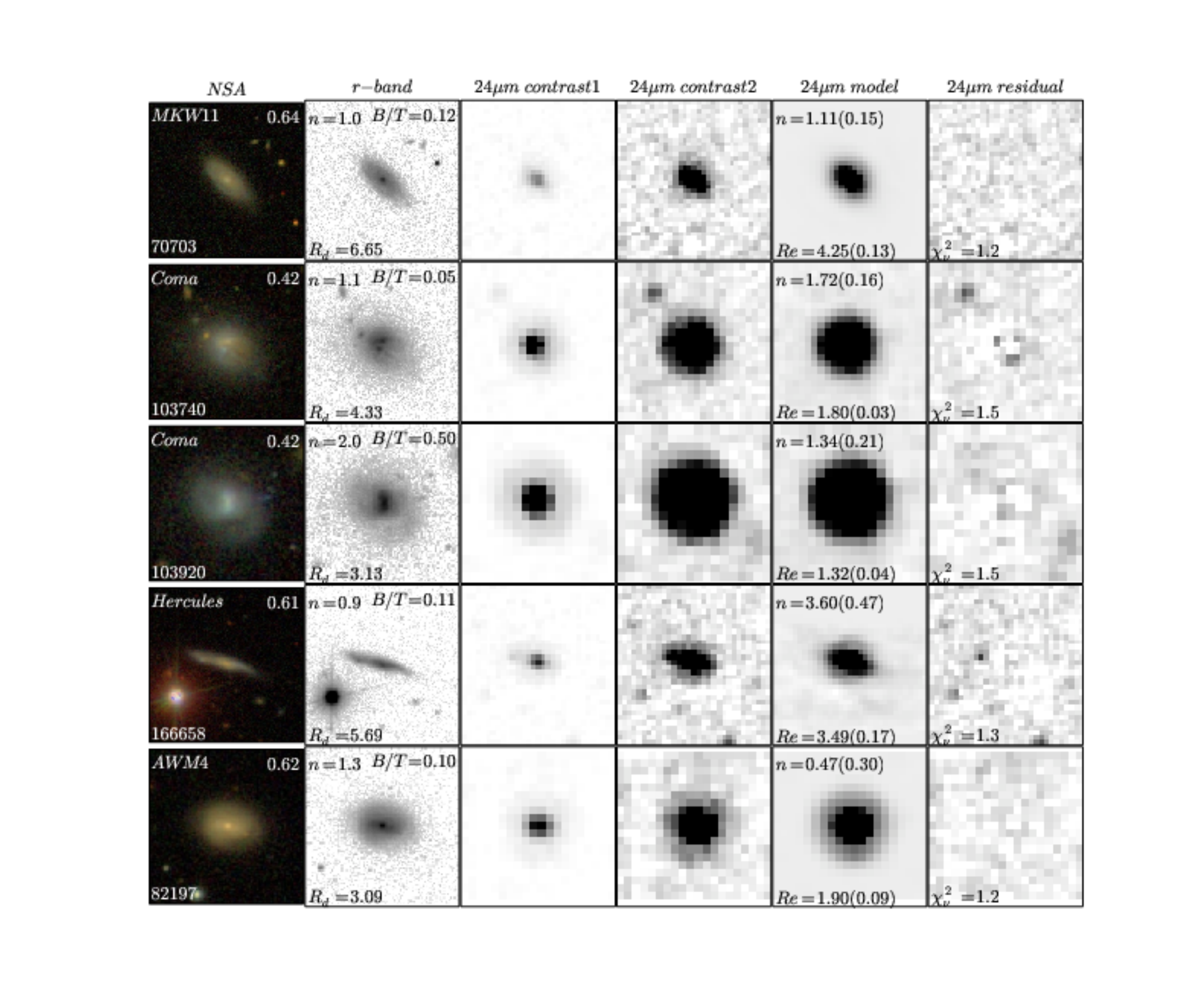}
\caption{\small Example GALFIT modeling results for
  five randomly selected galaxies with\sized \ between 0.4 and 0.7. Columns 1 and 2 show the
  SDSS color and $r$-band images.  Columns 3 and 4 show the 24\micron
  \ image at two different stretches to emphasize the high and low
  surface-brightness features, respectively.  Columns 5 and 6 show the
  24\micron \ single-component \sers \ model
  and residual, and the contrast is identical to column 4 to emphasize the low
  surface-brightness features.  The text in column 1 gives the parent cluster, the NSA ID, and
\sized. Column 2 lists the $r$-band \sers \ index n, the
bulge-to-total ratio, and the disk scale length in arcseconds.  The text in column 5
lists the 24\micron \ \sers \ index and effective radius in arcseconds, and column 6
lists $\chi^2_\nu$ of the best-fit model.}
\label{medsize}
\end{figure*}

\begin{figure*}[h]

\includegraphics[width=\textwidth]{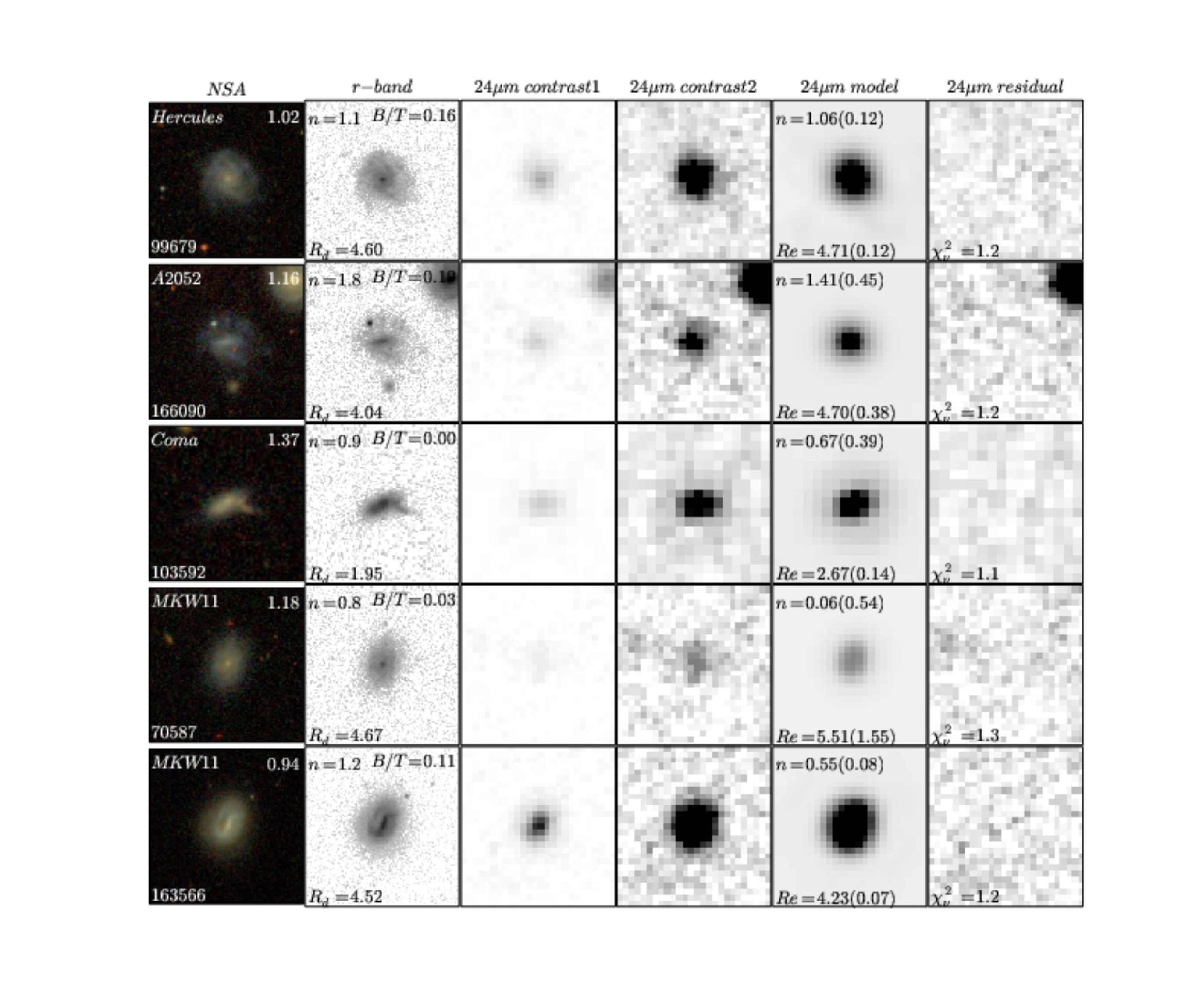}
\caption{\small Example GALFIT modeling results for
  five randomly selected galaxies with
  \sized \ greater than 0.9.  Columns 1 and 2 show the
  SDSS color and $r$-band images.  Columns 3 and 4 show the 24\micron
  \ image at two different stretches to emphasize the high and low
  surface-brightness features, respectively.  Columns 5 and 6 show the
  24\micron \ single-component \sers \ model
  and residual, and the contrast is identical to column 4 to emphasize the low
  surface-brightness features.  The text in column 1 gives the parent cluster, the NSA ID, and
\sized. Column 2 lists the $r$-band \sers \ index n, the
bulge-to-total ratio, and the disk scale length in arcseconds.  The text in column 5
lists the 24\micron \ \sers \ index and effective radius in arcseconds, and column 6
lists $\chi^2_\nu$ of the best-fit model.}
\label{largesize}
\end{figure*}

We show examples of the 24\micron \ \sers \ models in Figures 
\ref{smallsize}$-$\ref{largesize}.  In Figure \ref{smallsize}, we
show five randomly selected galaxies whose 24\micron \ emission is
more compact than the $r$-band emission, specifically $0.1 < R_{24}/R_d <
0.3$.  Columns 1 and 2 show the
  SDSS color and $r$-band images, respectively.  Columns 3 and 4 show the 24\micron
  \ image at two different stretches.  The first stretch ($contrast1$)
  highlights the high surface-brightness features, and the second
  stretch ($contrast2$) emphasizes the low surface-brightness features.  Columns 5
  and 6 show the 24\micron \ \sers \ model
  and residual, and both are shown at $contrast2$.  Figure \ref{medsize} shows five randomly selected
  galaxies with  $0.4 < R_{24}/R_d < 0.7$, and Figure
  \ref{largesize} shows five randomly selected galaxies with
 $R_{24}/R_d > 0.9$.  The columns are the same as for Figure
  \ref{smallsize}.


\begin{figure*}[h]
\includegraphics[width=.7\textwidth]{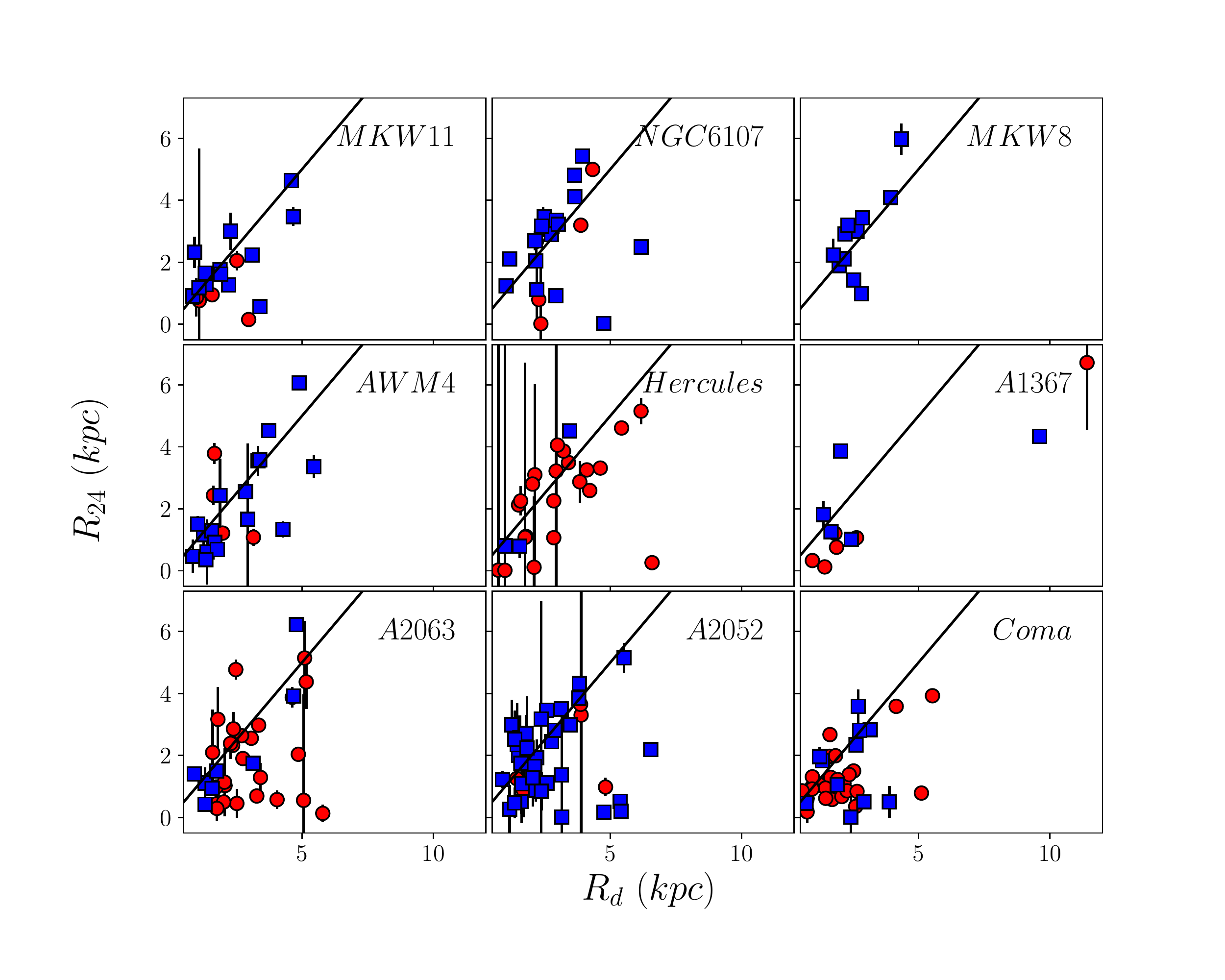}
\caption{\small $R_{24}$ versus $R_e(r)$ for all \lcs \ clusters.  The clusters are 
ordered left-to-right and top-to-bottom in terms of increasing X-ray luminosity.  
The solid black line shows a one-to-one correlation.  
The red circles and blue squares show
the core and external galaxies, respectively.  
For all 
clusters, the $R_{24}$ is systematically smaller
than $R_d$.   }
\label{re24vsre}
\end{figure*}

In Figure \ref{re24vsre}, we plot
the effective radius of the best-fit 24\micron \ model versus the 
$r$-band half-light radius for the disk component from
\citet{simard11} 
for each cluster pointing.  The clusters are 
ordered left-to-right and top-to-bottom by increasing X-ray luminosity.  
The solid black line shows a one-to-one correlation.
The circles and squares show core and external galaxies,
respectively.  For all 
clusters, the 24\micron \ and $r$-band sizes are correlated (see also
Figure \ref{GALFITvsNSA}), 
although the 24\micron \ half-light radii are systematically smaller
than the $r$-band disk half-light radii $R_d$.  

\begin{figure*}[h]
\plottwo{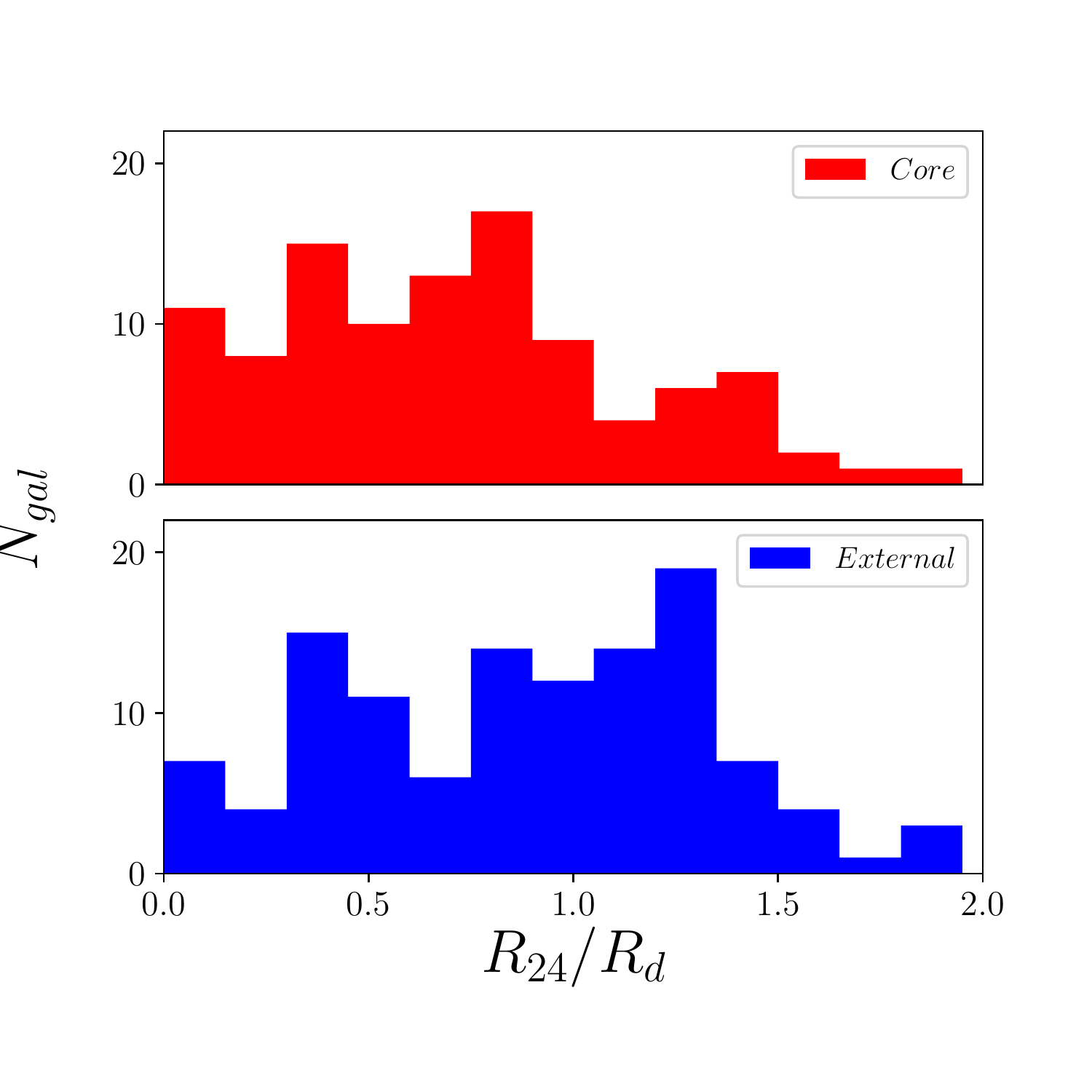}{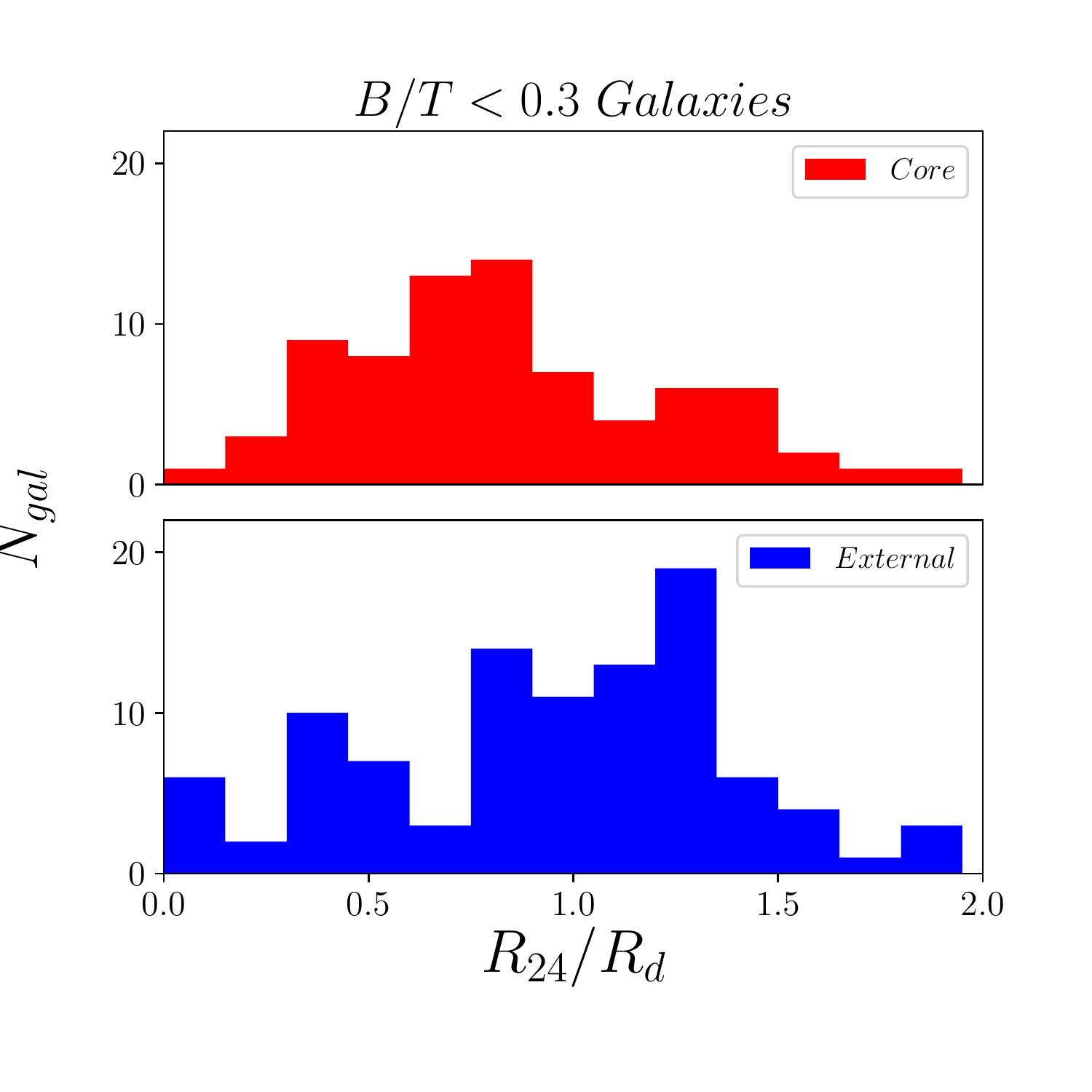}
\caption{\small (Left) Distribution of \sized \ for external (bottom;
  blue histogram) and core
  (top; red histogram)  galaxies.  The distribution of \sized \ for core galaxies
  is offset to smaller values compared to the distribution for the external
  galaxies. (Right) Same as left panel but for $B/T < 0.3$ galaxies
  only to help control for dependence of \sized \ on $B/T$.  Again, the core galaxies are offset to smaller values of \sized.
\label{sizehist}}
\end{figure*}

We show the distribution of \sized \ for the core and external 
galaxies in Figure \ref{sizehist}.  The
mean (median) size ratio for the core galaxies is $0.73 (0.69) \pm
0.04$, 
where the uncertainty is the error in the mean.  
The mean (median) size ratio for the external galaxies is $0.91 (0.94)
\pm 0.04$. 
We use both the K-S
and Anderson-Darling tests to compare the distribution of \sized \ for
the core and
external galaxies.  Both tests reject the null hypothesis at the
{$3\sigma$} level, which indicates that the core galaxies have
significantly smaller values of \sized.
We therefore find that the spatial distribution of star-formation in core galaxies
is more concentrated than in external galaxies.

\begin{table}
\caption{Definition of Radial Size Symbols}
\label{radii}
\begin{tabular}{p{.5in}p{2.6in}}
\hline \hline
Symbol & Definition \\
\hline
$R_{24}$ & 24\micron \ half-light radius from GALFIT single-component
\sers \ model.\\ 
$R_{e}$  & $r$-band half-light radius from GIM2D single-component \sers \ model
\citep{simard11}.\\
$R_{d}$ & $r$-band half-light radius of disk component based on
two-component GIM2D model with $n=4$ bulge $+$ exponential disk \citep{simard11}.  \\
\hline
\end{tabular}
\end{table}

Numerous other studies have measured the relative size of the star-forming 
and stellar disks as a function of
environment at both low \citep{moss00, dale01, koopmann06b,schaefer17} and intermediate
redshift \citep{bamford07, bosch13}, and most find that
the star-forming region is more concentrated
among cluster galaxies than the field.  
\citet{jaffe11}
find that the outer extent of the emission-line region is
systematically smaller in cluster galaxies. 

Our quantitative size estimates of \sized \ for core
(mean, median, error in the mean: 0.73, 0.69, 0.04) and external (mean, median, error in the mean: 0.91, 0.94, 0.04)
galaxies are in reasonable agreement with previous
work. \citet{koopmann06b} find $r_{H\alpha}/r_R =  0.91 \pm 0.05$ and $1.18\pm0.10$ for
Virgo cluster and field galaxies, respectively (error is error in the mean).  
At higher redshift, \citet{bosch13} find $\sim0.9$ and 1.27 for cluster and field galaxies
at $z=0.2$, and \citet{bamford07} find
$r_{\rm em}/r_B = 0.92\pm 0.07$ and $1.22 \pm 0.06$ for $0.2 \lesssim z
\lesssim 0.8$ cluster and field galaxies.  \citet{jaffe11}
find a ratio of $\sim 0.8$ for $0.4 < z < 1$ galaxies that is
independent of environment.  Interestingly, \citet{jaffe11}
use the ESO Distant Cluster Survey for their cluster sample, and these
are relatively low-mass clusters.
With the exception of \citet{jaffe11}, our values are systematically 
lower than most previous 
measurements that have been made using optical emission lines, 
but the $\sim20$\% offset we find between field and cluster sizes
is consistent with previous work.

\subsection{\sized \ versus $B/T$, Local Galaxy
  Density, and Stellar Mass}
\label{sizestuff}

\begin{figure*}[h]
\includegraphics[width=.95\textwidth]{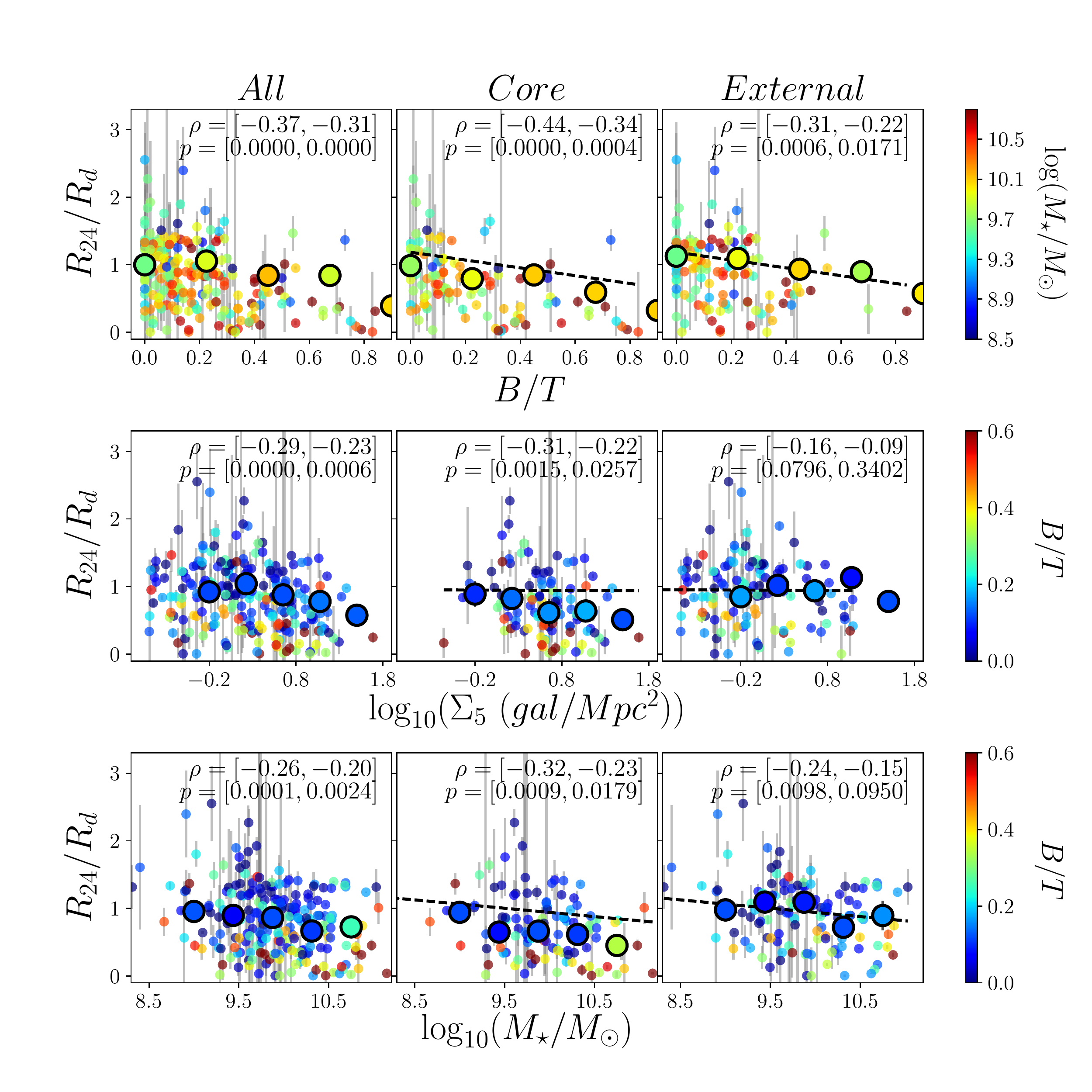}
\caption{\small (Top) \sized \ versus
$B/T$  for all (left), core (center), and external galaxies
(right).  Points are colored according
to stellar mass.  The 68\% confidence interval for the Spearman rank correlation coefficient, $\rho$, and
the corresponding $p-value$ are shown in the upper right.  
The large points show the median in five equally populated
bins.  The black dashed line in the center and right columns is a
simple linear fit to the external sample.  We show it with the core
panel to aid in comparing the samples.  (Middle) \sized \ versus $\Sigma_5$.  Points are colored
according to $B/T$.  \sized \ and $\Sigma_5$ are strongly correlated
for the full sample.  (Bottom) \sized \ versus $\log_{10}(M_*/M_\odot)$, the
local galaxy density. The external sample shows a significant
anti-correlation between \sized \ and stellar mass, whereas the core
galaxies show a flat relation.}
\label{size3column}
\end{figure*}

We show how \sized \ varies with {bulge-to-total ratio} in the top row of
Figure \ref{size3column}.  
As mentioned in Section \ref{selection}, the $B/T$ values are from \citet{simard11}.
The sample is divided
into full (left), core (middle), and external (right).  The points are colored according to
stellar mass.  
The Spearman rank coefficient and
probability of the null hypothesis are shown in the top of each panel.  
To account for the variation in error among the
individual 24\micron  \ best-fit models, we create 1000 Monte-Carlo realizations of the data
and calculate the Spearman rank correlation coefficient for each
realization, assuming that the GALFIT errors for $R_{24}$ are
normally distributed.  We report the 68\% confidence
interval for the correlation coefficient ($\rho$) and the probability of
no correlation ($p$). 
The $p$-values indicate that \sized \ and $B/T$ are
inversely correlated and that the correlation is strong.

{We use bulge-to-total ratio instead of
  visually classified morphology because \citet{koopmann98} show that
  visual classification is biased by the specific star-formation
  rate.  More specifically, when comparing galaxies of a fixed
  concentration, \citet{koopmann98} show that galaxies of a with low star-formation rates are systematically
  classified as having an earlier Hubble type. }

The \sers\ index of a single-component fit can also be used as a proxy for
morphology, with the index increasing from 1 to 4 as a galaxy profile
goes from disk dominated to bulge dominated.  
We find a similar correlation between \sized \
versus the $r$-band \sers \ index, which is not surprising given that
$B/T$ and \sers \ index are themselves strongly correlated. 
The result, then, is that galaxies with more concentrated stellar profiles have
smaller \sized, and at a given $B/T$, core galaxies have smaller
\sized \ ratios than the external galaxies.  

In the middle panel of Figure \ref{size3column}, we show \sized \
versus local galaxy density.  
\sized \ and $\Sigma_5$ are strongly
correlated as indicated by a Spearman rank test.
A similar effect can be seen in Figure \ref{phasespace}; galaxies with smaller \sized
\ ratios are more likely to be found at low projected radii where the
projected density of galaxies is high.

The correlation between the radial distribution of star formation and
morphology has been known for a long time. 
Mapping the distribution of H~II regions in 37 nearby galaxies,
\citet{hodge83} find that
early-type spirals tend to have centrally concentrated, symmetric star formation
whereas later-type spirals have more extended and asymmetric emission.
The results were confirmed by \citet{bendo07}, who use 24\micron \ emission to quantify the spatial distribution of star formation for the 65
galaxies in the {\it Spitzer Infrared Nearby Galaxy Survey}.  
However, not all previous studies find a link between the spatial
distribution of star formation and galaxy morphology \citep{dale01,koopmann06b,fossati13}.  
Some of these discrepancies might be due to differences between
samples.  For example, the \citet{dale01} sample
is dominated by late-type spirals, and the galaxies were selected to
have strong emission-line features; both of these 
selection effects could hinder their ability to detect a trend with
morphology.  
\citet{koopmann06b} do not use the central
region of each galaxy when fitting the \ha \ radial profiles, due to
possible bulge contamination and extinction problems.  
Thus no galaxies with weak or severely truncated star formation have
measured scale lengths.  
Interestingly, \citet{bretherton13} find a correlation between the relative size of
the star-forming disk and morphology among field galaxies but not among cluster
galaxies.  
However, we find a correlation between \sized \ and
$B/T$ for both core and external galaxies.

Comparison of the core and external panels shows that the core
galaxies have lower values of \sized.  
To further emphasize the offset and to control for the correlation
between morphology and environment, we again show the distribution of
relative sizes for both core and external galaxies but for $B/T < 0.3$
galaxies only in the right panel of Figure
\ref{sizehist}.  The core galaxies have \sized \ ratios that are clearly offset to
smaller values.

In the bottom row of Figure \ref{size3column}, we show
\sized \ versus stellar mass, where  
 stellar mass is calculated as described in 
\citet{moustakas13} and assumes a Chabrier IMF \citep{chabrier03}.  
The left column shows the entire GALFIT
sample, and the middle and right columns show the core and external samples separately.   
Using the full sample, we test for a correlation between stellar mass and \sized \ using a
Spearman rank test.  
The results indicate that \sized \ and stellar mass are inversely
correlated, although the correlation is not particularly strong.

\begin{figure}[h]
\includegraphics[width=.5\textwidth]{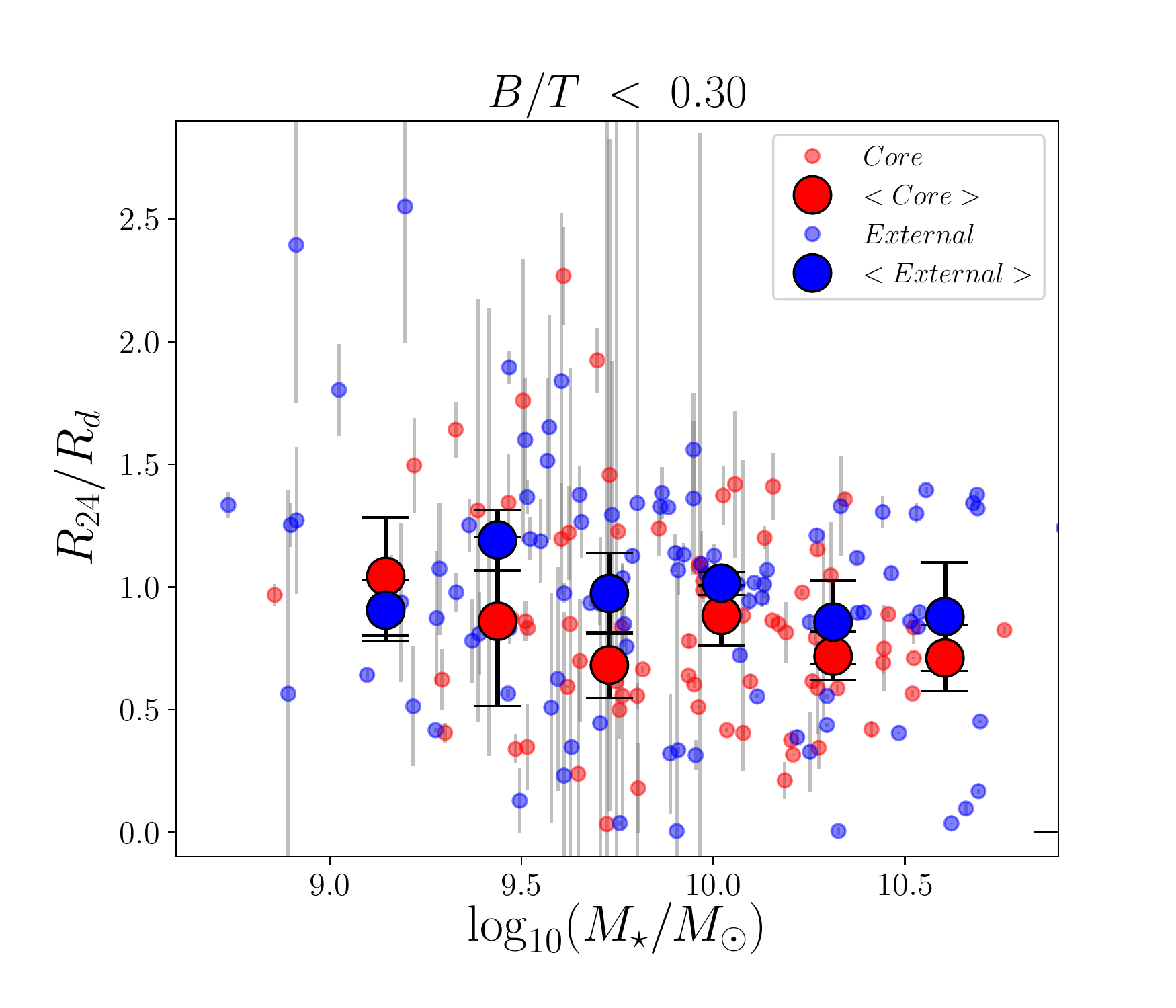}
\caption{\sized \ versus stellar mass for core (red) and external
  (blue) galaxies.  We include $B/T < 0.3$ galaxies only to help limit
  any systematics with $B/T$.  The small points show \sized \ for individual
  galaxies, and the large circles show the median in equally spaced
  bins.  The error on the binned points is the 68\% confidence
  interval, which we calculate using bootstrap resampling of the
  galaxies in each bin.  While \sized \ is systematically lower for
  the core galaxies, the difference between core and external size
  ratios does not appear to depend on stellar mass.}
\label{sizestellarmass}
\end{figure}

We investigate the dependence of \sized \ on stellar mass further in
Figure \ref{sizestellarmass}.  To control for morphology, we include
only the $B/T < 0.3$ galaxies.   We show the core galaxies in red and
the external galaxies in blue.  The large red and blue circles show
the median \sized \ for the core and external samples, respectively,
in equally spaced bins.  
The errorbars show the 68\% confidence interval on the median, 
which we calculate using bootstrap resampling of the galaxies in each bin.
While $R_{24}/R_d$ is systematically lower for the core galaxies, the
difference between the 
core and external size ratios does not appear to depend on stellar mass.

We also look at \sized \ versus several combinations of stellar mass
density $--$ namely, stellar mass surface density 
\citep{zhang13}, and $M_*/R_e$ \citep[e.g.][]{omand14}.  We 
detect no correlation between \sized \ and $M_*/R_e^2$, and a moderate
correlation between \sized \ and $M_*/R_e$.  However, we are testing
for a correlation between two correlated variables because both \sized \ and 
$M_*/R_e$ have a measure of $r$-band size in the denominator, and the significance of
any detected correlation is thus difficult to interpret.  

\subsection{Partial Correlation Analysis}
The previous section provides convincing evidence that \sized \
depends on $B/T$, environment, and stellar mass.  However, $B/T$,
environment, and stellar mass are correlated with each other.  We attempt to
separate the influence of these variables using partial correlation analysis.
The partial correlation statistic, which we compute
using the ppcor.pcor package in $R$, indicates the degree to which Y
is correlated with X, after removing any correlation between Y and other
variables W and Z.  The partial
correlation coefficient, $\rho$, can range between $-1$ and 1 for
strongly anti-correlated and correlated variables, respectively.   
We again account for the error in \sized \ by creating 1000 realizations of
our data, and  the mean and 68\% confidence
interval for the correlation coefficient 
and the corresponding corresponding statistic are shown in Table
\ref{pcor}.  The statistic indicates the significance
of the corresponding correlation (e.g. $3\sigma$).
We highlight 
the significant correlations in bold.  

The results
indicate that \sized \ is correlated with $\Sigma_5$ even after
controlling for variations in \sized \ with $B/T$ and stellar
mass. Similarly, \sized \ is correlated with $B/T$ even after
controlling for any variations of \sized \ with $\Sigma_5$ 
and $M_*$.  However, both the Spearman rank
and partial correlation tests indicate that \sized \ is more strongly
correlated with $B/T$ than with $\Sigma_5$.  Interestingly, the
partial correlation analysis indicates that \sized \ is not correlated
with stellar mass after variations with $B/T$ and $\Sigma_5$ are
removed.  

\subsection{Impact of Coma}
\label{comaresults}
Coma is the richest, most X-ray luminous cluster in our sample, and in
this section we briefly discuss how the properties of Coma galaxies
affect our results.  First, we recompute the partial correlation coefficients
of \sized, $M_*$, $B/T$, and $\Sigma_5$ after removing Coma galaxies
from our sample.  We find that the correlation between \sized \ and
$B/T$ is still significant at the $4.0\sigma$ confidence level; $B/T$
remains the most strongly correlated variable with \sized.  We detect a
less significant ($1.9 \sigma$) correlation between \sized \ and
$\Sigma_5$.   \sized \ and $M_*$ remain 
uncorrelated when Coma is removed.
The results of the partial correlation analysis are
summarized in the last two columns of Table \ref{pcor}.

Second, we also compare with distribution of \sized \ for $B/T < 0.3$
core and external galaxies.  The mean (median) \sized \ for the core and external galaxies is
$0.74 (0.72) \pm 0.05$  and $0.93 (0.95) \pm 0.05$, respectively, and a KS test indicates that the core
and external distributions differ at the $3\sigma$ level.  Thus, even though the
correlation between \sized \ and $\Sigma_5$ weakens when Coma is
removed, we still find that the core galaxies have significantly
smaller \sized \ ratios than the external galaxies.  


\begin{table*}[h]
\caption{Summary of Partial Correlation Analysis}
\label{pcor}
\begin{center}
\begin{tabular}{l|cccc|cccc}
\hline \hline
& \multicolumn{4}{|c}{With Coma ($N = 224$) } &
\multicolumn{4}{|c}{Without  Coma ($N = 192$)} \\
 & \multicolumn{1}{c}{$\rho$} & Conf.&
 \multicolumn{1}{c}{Significance} & Conf.&\multicolumn{1}{|c}{$\rho$} & Conf.&
 \multicolumn{1}{c}{Significance} & Conf.\\
& & Interval & &Interval & &Interval & &Interval \\
\hline 

\sized $-$$M_*$
& $-$0.16 & [$-$0.19, $-$0.12]  & 2.4 &  [2.9, 1.9] 
&$-$0.13 &[$-$0.17, $-$0.09] & 1.9 & [2.4, 1.3]   \\

\sized $-$$B/T $          
& {\bf $-$0.29} & [$-0.32, -0.26$]&{\bf4.5}& [5.1, 3.9] 
& {\bf $-$0.28}& [$-$0.32, $-$0.24]&{\bf 4.0} & [4.6, 3.3] \\

\sized $-$$\Sigma_5 $ 
&{\bf $-$0.24}& [$-0.28, -0.21$] & {\bf 3.7}&[4.3, 3.2]  
&$-$0.17 & [$-$0.21, $-$0.13] & 2.4 &[2.9, 1.8] \\

\hline
\end{tabular}
\end{center}
\end{table*}

\subsection{\sized \ versus Large-scale Environment}
\begin{figure}[h]
\includegraphics[width=.5\textwidth]{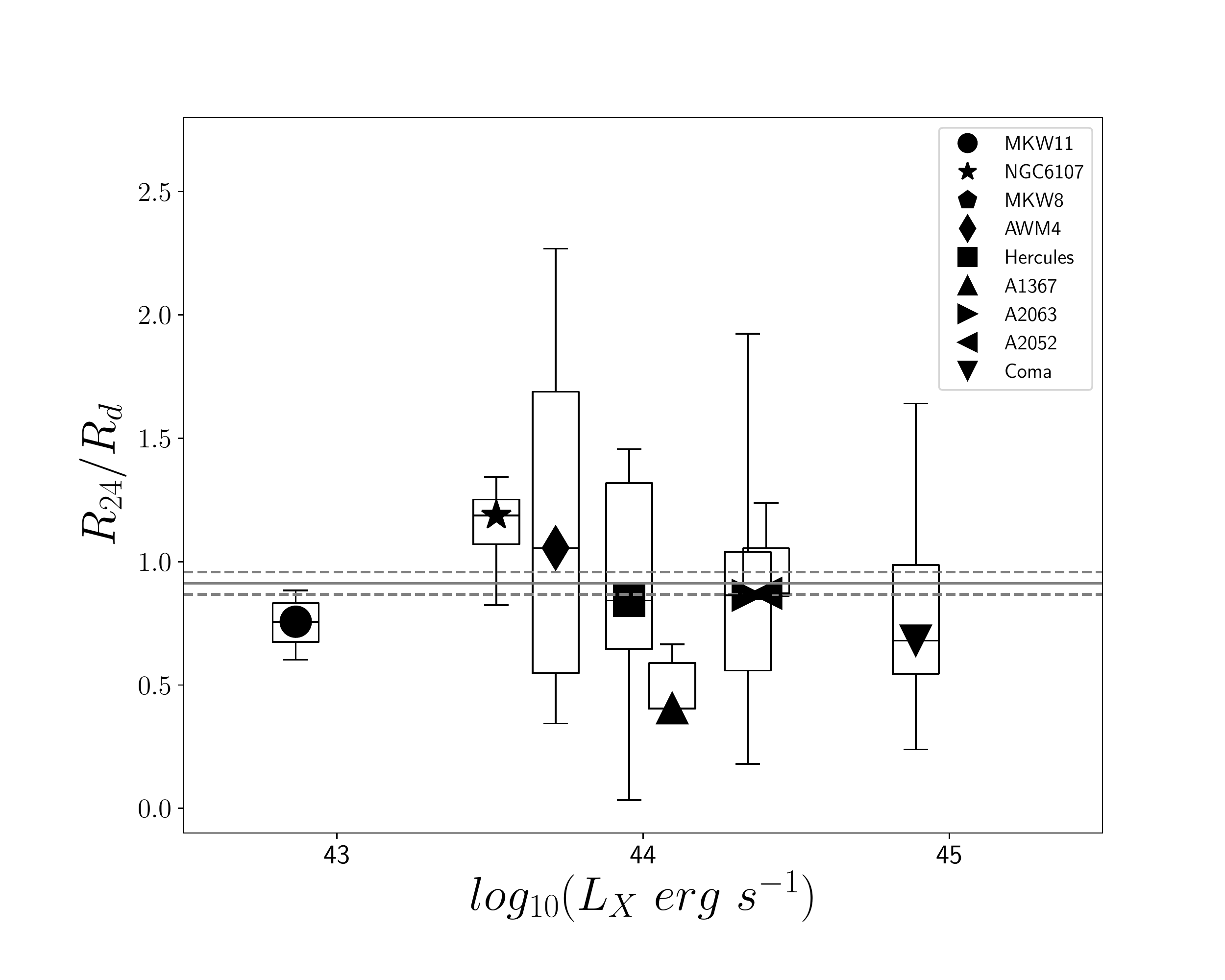}
\caption{\small Average \sized \ of $B/T < 0.3$ galaxies versus the X-ray luminosity of
  cluster.  The individual points show
the median value, the box marks the range between the upper and
lower quartiles, and the whiskers show the full range of values for
each group/cluster.  The horizontal line shows the
average \sized \ for the field galaxies, and the dashed
lines show the error in the mean.  }
\label{sizelx}
\end{figure}

The density of the intra-cluster medium is a key factor in determining the effectiveness
with which gas is removed from infalling galaxies by ram-pressure stripping \citep[e.g.][]{gunn72}. 
We do not have a measure of $\rho_{\rm ICM}$, so we use X-ray luminosity as
a proxy.  
We show \sized \ for $B/T < 0.3$ cluster galaxies versus the
X-ray luminosity of each cluster in Figure \ref{sizelx}.  Again, we
limit the range of $B/T$ in this comparison to help control for the
correlation between $B/T$ and environment.  We use a box and whisker plot to
show the range of values for each cluster.  The box extends from the
lower to upper quartiles, while the whiskers (errorbars) show the range
in each bin.
The horizontal line shows the
average \sized \ for the external sample, and the dashed
lines show the error in the mean.  
The MIPS scans for A1367, Hercules, and Coma do not extend to \rtwo.
We do not attempt to correct for this, however, because if
we try to match the areal coverage of these clusters by selecting a
smaller radial cut, we run out of galaxies in other clusters.  
We repeat the same analysis using X-ray temperature in place of X-ray
luminosity, and we find similar results.

\begin{figure}[h]
\includegraphics[width=.5\textwidth]{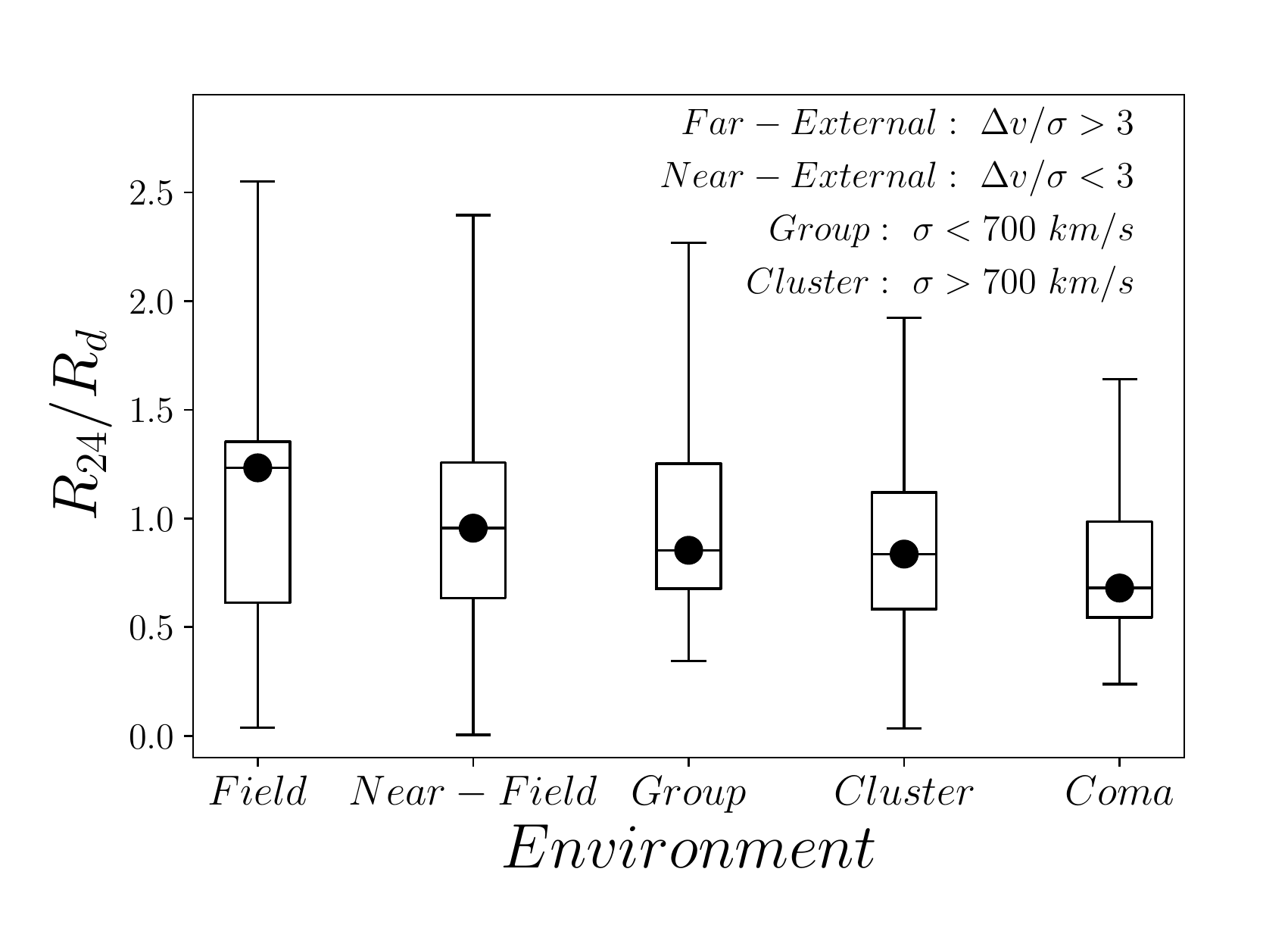}
\caption{\small Median \sized \ of $B/T < 0.3$ galaxies versus
  large-scale environment.  The boxes show first and third quartiles,
  and the whiskers or errorbars show the range of the data in each bin.  Here we split the external sample into
  near-external ($\Delta v/\sigma < 3$) and the far-external ($\Delta
  v/\sigma > 3$).  We divide the clusters by velocity dispersion,
  defining groups as $\sigma < 700~\rm km~s^{-1}$.  We leave Coma in a class by
  itself.  }
\label{size_env}
\end{figure}

Due to the relatively small numbers of galaxies in each cluster, we next
look for trends with the large-scale environment by binning the clusters according to X-ray luminosity or velocity
dispersion.  We combine the galaxy groups ($\sigma < 700$~km~s$^{-1}$: MKW11,
MKW8, AWM4, and NGC6107) and clusters ($\sigma > 700$~km~s$^{-1}$: Hercules,
Abell~1367, Abell~2052, and Abell~2063), and leave Coma in a class by
itself.  For this comparison only, we separate the external sample into
near-external and far-external galaxies, where the near-external galaxies have
$\Delta v/\sigma < 3$.  These galaxies likely live
in the large-scale structure surrounding the clusters and may already
be affected by environmental process. The far-external
galaxies have $\Delta v/\sigma > 3$ and are more likely to be isolated from
the cluster environment.  We show the median \sized \ for $B/T < 0.3$
galaxies versus global environment in Figure \ref{size_env}.  Again,
we use a box and whisker plot to show the inner quartiles and range of
the data.  The median \sized \ decreases as the environmental density
increases from the field through to Coma.   In addition, the galaxy
groups and lower-mass clusters have median \sized \ that falls between
the near-field and Coma,
but the differences are not statistically significant.  
While our results suggest that cluster mass is important, a larger
sample of groups and clusters is needed to explore variations between
\sized \ and cluster/group mass.
Not many studies have measured \sized \ as a function of cluster mass
or X-ray luminosity.
\citet{dale01} make a preliminary attempt to measure the effect of
cluster X-ray luminosity on stripping, and they find no systematic
trend.

 \subsection{\sized \ versus H~I Content} 
\begin{figure*}[h]
\plottwo {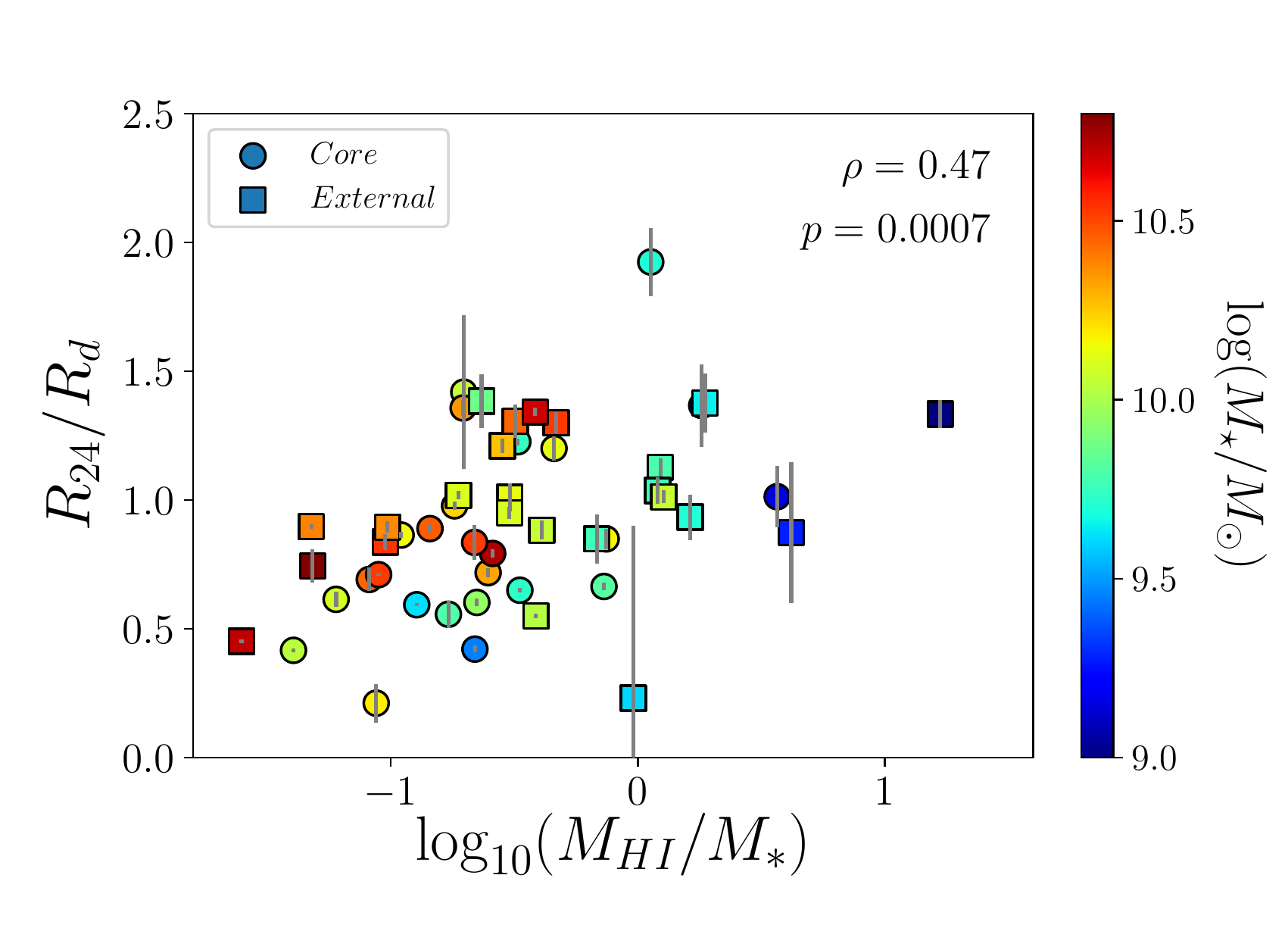}{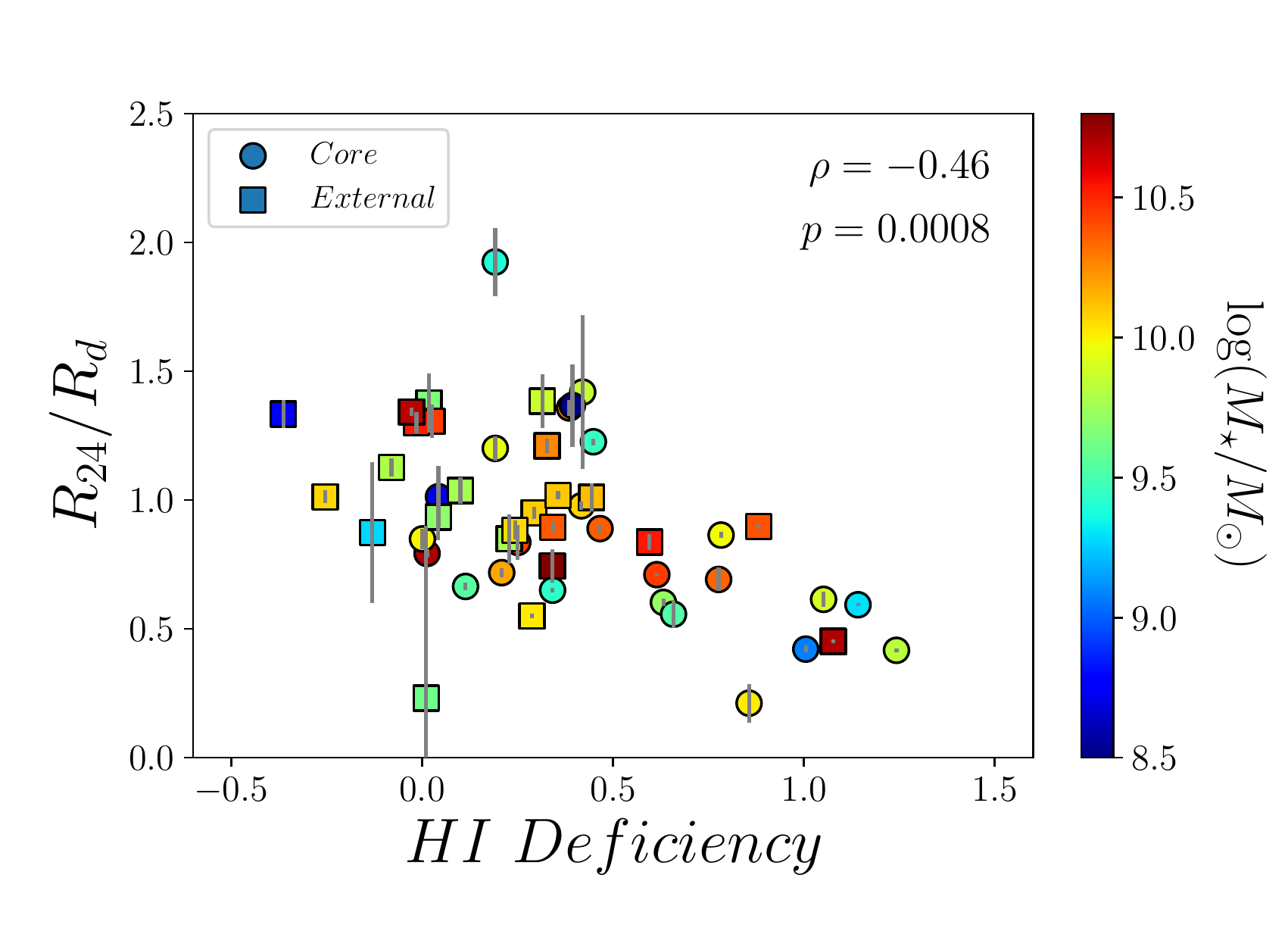}
\caption{(Left) \small \sized \ versus
H~I mass fraction.  
The data show a significant ($>3\sigma$) correlation
between \sized \ and H~I mass fraction. 
(Right) \sized \ versus
H~I deficiency.  The data show a significant ($>3\sigma$) anti-correlation
between \sized \ and H~I deficiency in the sense that galaxies with smaller \sized \ also have
less H~I gas than isolated galaxies.
\label{sizeHIdef}}
\end{figure*}

All of the clusters in our sample lie within the ALFALFA survey \citep{giovanelli05}.  
ALFALFA maps H~I to a resolution of 3.\arcmin5 and with a pointing accuracy
better than 30\arcsec \ for sources with $\rm S/N > 6.5$, 
and a total of 43 galaxies in the GALFIT sample have H~I detections. 
We calculate H~I mass for these galaxies according to the following relation:
\begin{equation}
M_{H~I} = 2.36 \times 10^5 \ D^2 \int F dV \ M_\odot,
\end{equation}
where $D$ is the distance in Mpc and $\int F dV$ is the total H~I line
flux in Jy-km~s$^{-1}$ \citep[e.g.][]{wild52,roberts62}.
In the left panel of Figure \ref{sizeHIdef} we
show \sized \ versus H~I mass fraction
$M_{H~I} / M_*$.  The squares and circles show external and core
galaxies, respectivly, and the color indicates stellar mass.   A
Spearman rank test indicates that the two quantities are strongly
correlated;
galaxies with smaller values of \sized \ have a smaller
H~I mass fraction.

Another common way to describe the gas content of spiral galaxies is with H~I
deficiency 
\citep[][]{haynes84,toribio11}.  
This compares the H~I content of a galaxy to the H~I content of isolated
field galaxies of comparable size, and positive values of deficiency indicate that a galaxy has
{\em less} H~I gas than a field galaxy of comparable size \citep{haynes84, toribio11}.    
We calculate H~I deficiency according to the relationship presented in
\citet{toribio11},  using the isophotal $r$-band 
diameter from SDSS to measure size.
In Figure \ref{sizeHIdef} we show \sized \ versus H~I deficiency.
 The data show a significant ($3.5\sigma$) anti-correlation
 in the sense that galaxies with smaller \sized \ also have
 less H~I gas than isolated galaxies of comparable size.  
We are not able to confirm that the converse (i.e., that H~I deficient galaxies will have small \sized \ ratios) is true; 
H~I deficient galaxies with widespread but low
 levels of star-formation would likely fall below our
 surface-brightness cut.  
Figure
\ref{sizeHIdef} also shows that core galaxies tend to have lower H~I mass
fractions and higher H~I deficiencies than external galaxies.  

Our findings are consistent with previous studies.  The $H\alpha 3$ survey of nearby galaxies \citep{gavazzi13,
  fossati13} and \citet{koopmann04b} for Virgo galaxies find that the
relative size of the \ha \ disk is inversely correlated with H~I deficiency.  
In addition, infrared and CO observations of Virgo galaxies show that H~I
deficient spirals have smaller dust and molecular gas disks \citep{cortese10,boselli11}.

\subsection{\sized \ versus Color}

\begin{figure}[h]
\includegraphics[width=.5\textwidth]{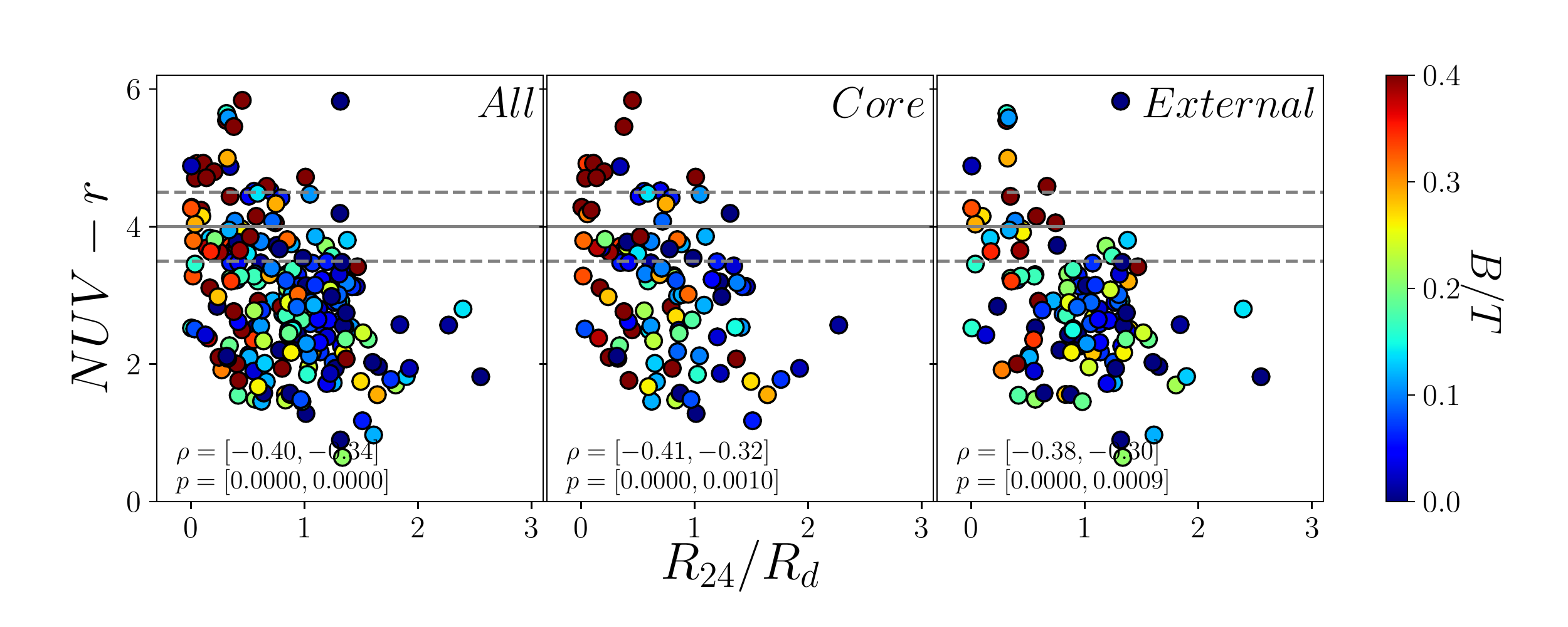}
\caption{\small NUV$-r$ versus \sized \ color for all (left panel), core (middle panel),
  and external (right panel) galaxies.  The points are colored by
  $B/T$.  The horizontal solid line in each panel marks ${\rm NUV}-r =
4$, a rough dividing line
between blue and red galaxies \citep[e.g.][]{salim07}, and the dashed
horizontal lines mark the region of the green valley
\citep[e.g.][]{wyder07}.  Galaxies with redder ${\rm NUV}-r$ colors have
smaller values of \sized.
  }

\label{NUVrsize}
\end{figure}

If Figure \ref{NUVrsize} we show ${\rm NUV}-r$ versus \sized, and the points are color-coded by $B/T$.
The core and external galaxies are shown in the middle and right panels,
respectively.  The horizontal solid line in each panel marks ${\rm NUV}-r =
4$, a rough dividing line
between blue and red galaxies \citep[e.g.][]{salim07}, and the dashed
horizontal lines mark the region of the green valley
\citep[e.g.][]{wyder07}.  As before,  to account for the variation in error among the
individual 24\micron  \ best-fit models, we create 1000 Monte-Carlo realizations of the data
and calculate the Spearman rank correlation coefficient for each
realization, assuming that the GALFIT errors for $R_{24}$ are
normally distributed.  The numbers in Figure
\ref{NUVrsize} show the 68\% confidence
interval for the correlation coefficient ($\rho$) and the probability of
no correlation ($p$). 
\sized  \ is strongly inversely correlated with ${\rm NUV}-r$ color
($>5\sigma$ for the combined core$+$external sample); galaxies that
have red ${\rm NUV}-r$ colors and thus low NUV
specific star-formation rates have more centrally concentrated star
formation.  

Our results indicate that more concentrated
star formation is connected to a galaxy's transition from active to
passive.  Similarly, \citet{boselli14} find a correlation
between ${\rm NUV}-i$ color and molecular gas deficiency.  
In addition, the centrally concentrated star formation should lead to
an increase in $B/T$, as suggested by \citet{bosch13}.
We can
estimate the further evolution of the bulge-to-total ratio for the
green valley galaxies if we assume a typical gas mass fraction of 10\% (from Figure
\ref{sizeHIdef} for galaxies with \sized \ $\approx 0.5$).  If
we assume a current $B/T = 0.2$ and $M/L = 1$, the $B/T$ would increase to
0.3 after all the gas is consumed.  This is not a huge effect.
However, atomic gas is easier to strip than molecular gas
\citep[e.g.][]{boselli14}, and so our estimate is likely a lower limit
on the final bulge-to-total ratio.
We could better predict the growth of the central bulge if we obtain molecular gas masses for the \lcs \
galaxies.

\section{Discussion}
\label{discussion}

Our goal is to investigate the spatial distribution of star-formation
in galaxies that have been accreted into a cluster but are still
forming stars.  In Figures \ref{sizehist}-\ref{size_env}, we show that
star-forming galaxies in dense environments have more
centrally concentrated star formation.  
Models of environmentally driven depletion predict that gas is
preferentially removed from the outskirts of galaxies
\citep[e.g.][]{gunn72, kawata08,bekki14}, and our observations of
smaller star-forming disks in denser environments support this.  
In addition, we find that galaxies with
smaller star-forming disks are more H~I deficient
(Fig. \ref{sizeHIdef}) and have redder colors (Fig. \ref{NUVrsize}), 
suggesting that
smaller star-forming disks are indicative of the transition phase
between blue, gas-rich and red, depleted galaxies.  This transition is accompanied by
a modest increase in $B/T$.

\subsection{Outside-in Quenching Timescale}
The timescale over which \sized \ decreases in an important parameter
that can help identify the physical mechanism that is causing outside-in quenching.  
We construct a simple model to constrain this timescale using the
observed distribution of \sized \  for the core and external samples and
simulations of cluster infall.  Our model assumes that the size of the
star-forming disk decreases linearly with time once a galaxy is
accreted into a cluster.  The distribution of \sized \ for the core
galaxies results from the modification of the external \sized \
distribution as follows:
\begin{equation}
\left(\frac{R_{24}}{R_d}\right)_{core} = \left(\frac{R_{24}}{Rd}\right)_{external} \times \frac{dr}{dt}
\times t_{infall},
\label{sim-eqn}
\end{equation}
where $t_{\rm infall}$ is the time since accretion into the cluster, and
${dr}/{dt}$ is the rate at which the size of the star-forming disk is
decreasing in the cluster environment.  We assume ${dr}/{dt}$ is
the same for all galaxies and that the distribution of \sized \ in
the external sample is comparable to that of the core galaxies at
the time of infall.
(While field galaxies at the time of infall likely had higher star-formation
rates due to the correlation of galaxy star-formation rates with
redshift \citep[e.g.][and references therein]{madau14}, there is no compelling
evidence to suggest that \sized \ is larger for intermediate redshift galaxies \citep[e.g.][]{nelson16}. )
For each galaxy in the external sample, we assign an infall time as $\alpha t_{max}$,
where $\alpha$ is a random number between zero and 1.  In effect, this
assumes that the accretion rate is uniform over the time $t_{max}$,
which is a reasonable approximation to theoretical mass accretion
histories for $t_{max}<5$~Gyr \citep[e.g.][]{neistein08}.  
To be conservative, we let $t_{max}$, the time period over which the star-forming core
galaxies have been accreted, range from 1 to 4~Gyr.  We note that both
the phase space distribution of the core galaxies
\citep[e.g.][]{oman13} and simulations of
the mass accretion history of clusters \citep[e.g.][]{mcgee09} suggest that
$t_{max} \simeq 3-4$~Gyr for the star-forming core galaxies.

\begin{figure}
\includegraphics[width=.5\textwidth]{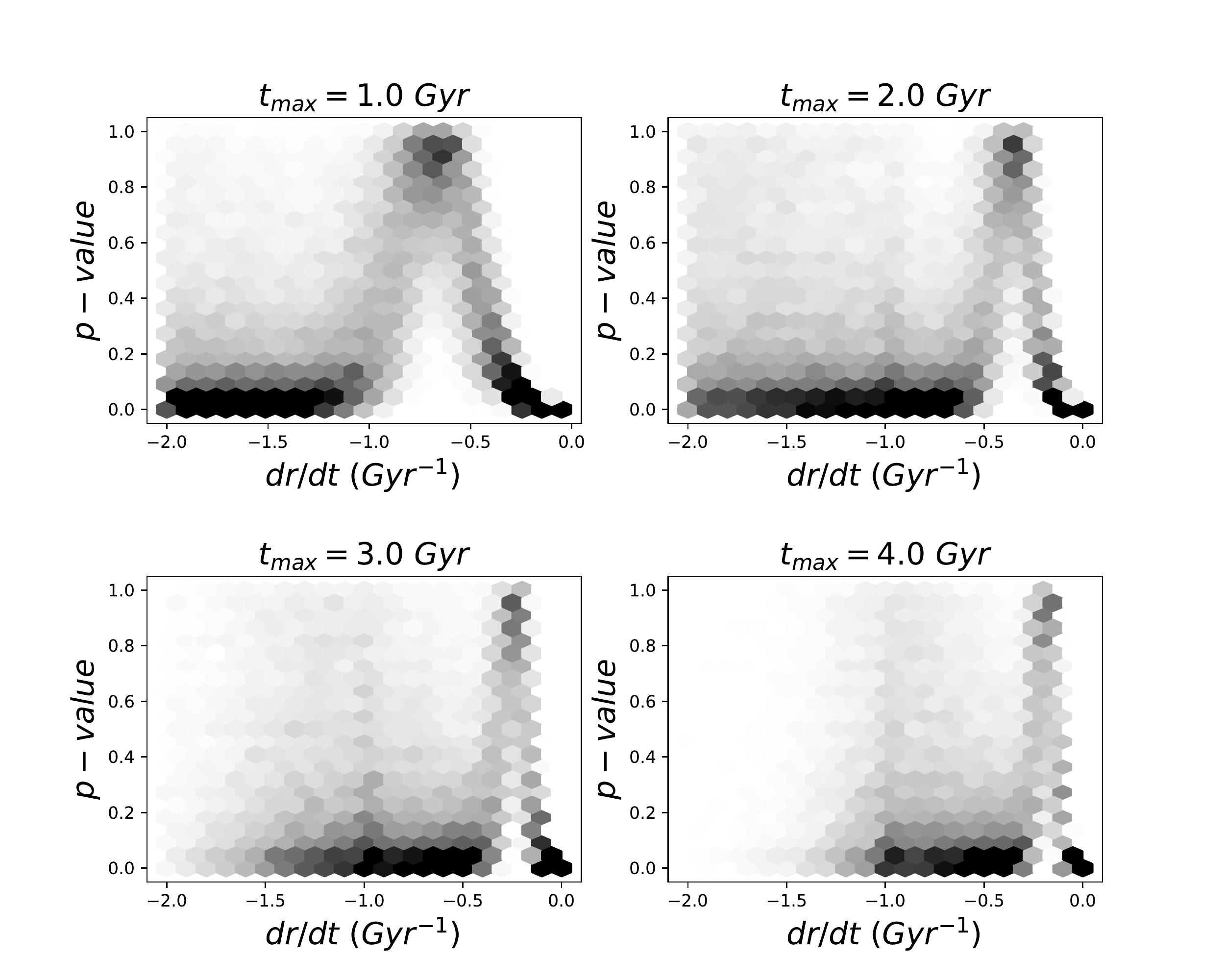}
\caption{We create a simple model where \sized \ of the external
  sample decreases uniformly at a rate of $dr/dt$ once a galaxy is
  accreted into a cluster.  We vary $dr/dt$
  from $-2$ to 0~Gyr$^{-1}$, and transform the \sized \ of the external sample by
  randomly assigning infall times to each galaxy (1000 realizations
  per $dr/dt$ value).  We compare the
  simulated distribution of \sized \ with the observed distribution of
  the core sample using a K-S test.  We show the distribution of K-S
  test probability values versus $dr/dt$.  The four panels show the
  results as we allow the accretion timescale for the core sample to
  increase from 1 to 4~Gyr.  For all accretion times, $dr/dt <
  1~Gyr^{-1}$, implying a long ($>1~$Gyr) timescale for removing the
  star-forming disk.}
\label{model}
\end{figure}

To quantify the timescale, we step $dr/dt$ from $-2$ to zero and
calculate the expected distribution of \sized \ for the core sample
according to equation \ref{sim-eqn}.  We compare
the resulting distribution of sizes to that of the observed core
sample, and compute the probability that the two are drawn from the
same population using a Kolmogorov-Smirnov test.  For a given value
of $dr/dt$, we repeat the comparison 1000 times, generating a new set
of infall times for each trial.  
We show the distribution of resulting K-S test $p$-values versus $dr/dt$ in
Figure \ref{model}, and $p$-values near 1 indicate that the simulated
distribution of \sized \ is similar to the observed distribution.
For $t_{max} = 1$~Gyr, $dr/dt$ peaks near 
$-0.6~Gyr^{-1}$, and the star-forming disk would be removed in $\sim
1.7$ Gyr.  For $t_{max} = 4$~Gyr, $dr/dt$ peaks near 
$-0.2~Gyr^{-1}$, and the star-forming disks would be completely removed in 5~Gyr.
Thus, while a precise estimate of the disk-shrinking time requires a
careful comparison with simulations of cluster growth, 
our simplified model suggests that for a reasonable estimate of infall
times, the disk-shrinking timescale is greater
than 1 Gyr and likely greater than 2~Gyr. 

This timescale has significant implications for the delay$+$rapid
quenching model of \citet{wetzel12, wetzel13}.  First, our estimated 
timescale is significantly longer than expected for the
“fast-quenching” phase of \citet{wetzel13}.  Second, we show that
galaxies
undergo a significant transformation during the delay phase, in
contrast with the delay$+$rapid model as originally proposed, in which
the galaxies remain unchanged in the period before the rapid quenching event.

\subsection{The Effect of $B/T$ on Quenching}
At this
point, we do not have a satisfactory explanation for why $B/T$
is inversely correlated with \sized.   
One mechanism that could explain the link between morphology and
star-formation properties of spirals is morphological quenching: the presence of
a bulge stabilizes the gas in a disk, and this prevents gravitational
collapse and star formation \citep[e.g.][]{martig09}.  However, it is not clear that
this would lead to more concentrated star formation, and many of our
galaxies with high $B/T$ were not detected in H~I, which implies that
they have lower gas masses rather than a reservoir of gas that never
formed stars.
Simulations provide conflicting results.  \citet{steinhauser12} and
\citet{mccarthy08} show that more concentrated galaxies are better
able to retain their gas in dense environments, which seems at odds
with our results, while \citet{jachym07}
show that galaxies with larger bulge fractions lose more
gas. 

One plausible explanation comes from \citet{solanes01}.  
In an extensive study of H~I in clusters, they find that
early-type spirals are more likely to be H~I deficient than Sbc-Sc
spirals.  The discrepancy persists out to 
projected radii of $\approx 4$~Mpc but not farther, which shows that
this is likely the result of an environmental rather than secular process.
\citet{solanes01} point out that the central H~I depressions observed
in some early-type spirals could amplify the effect of ram-pressure
stripping, as shown by \citet{moore99}.  This could imply that the trend
in \sized \ with morphology arises because high $B/T$ galaxies
are more susceptible to ram-pressure stripping.  Future simulations
may provide insight into how the bulge
fraction of a spiral affects its ability to retain gas in dense environments.

\section{Summary}
We present 24\micron \ size measurements for \ntot \ galaxies in nine
nearby galaxy groups and clusters.  We normalize the 24\micron \
effective radius by the disk scale length \citep{simard11} and look
for variations in this ratio as a function of morphology, environment,
and stellar mass.  Our
primary results are that (1) \sized \ is strongly
correlated with morphology for star-forming galaxies in the sense that galaxies with higher
bulge-to-total ratios or larger \sers \ indices have more
centrally concentrated star formation, and 
(2) star-forming galaxies in more dense environments have
more centrally concentrated star formation than galaxies in less dense
environments with similar
mass and $B/T$.  
Furthermore, we find that galaxies with smaller star-forming
disks tend to have lower H~I mass fractions and redder ${\rm NUV}-r$
colors, suggesting that at least some galaxies experience a decline in
\sized \ as they transition from blue to red.  

We do not detect any trend in the median \sized \
ratio of cluster galaxies versus X-ray luminosity of the host
galaxy cluster. However, when we bin our sample by environment and control for the morphology$-$density
relation by using only $B/T < 0.3$ galaxies, we do see a trend with large-scale environment that suggests
that \sized \ is highest in the field, lower in our low-mass clusters
and groups, and lower still in the extreme environment within the Coma cluster.  

We build a toy model to constrain the timescale over which the
star-forming disks shrink in the cluster environment.  When allowing
the galaxies to enter the cluster with a realistic range of infall
histories, we find that the star-forming disks in our sample will
shrink on a timescale longer than 1~Gyr and likely longer than 2~Gyr. 

Our results provide a new piece of information on what is
happening to galaxies after they have been accreted by a cluster but
while they are still able to form stars.  
In the context of recent hybrid models of
environmental quenching \citep[e.g.][]{wetzel12,balogh16}, our core
galaxies are considered to be in the delay phase, the period between being accreted
into the cluster and complete quenching of star formation.  
We present clear evidence that the spatial distribution of star-formation
becomes more concentrated during this delay phase.  Our results suggest that the quenching
timescale is long ($> 2$~Gyr), which is consistent with mechanisms
such as the removal of
halo gas through starvation \citep[e.g.][]{larson80} and the slow removal of cold disk gas
through extended ram-pressure stripping.

\acknowledgements
R.A.F. gratefully acknowledges support from NSF grant AST-0847430.
R.A.F., G.R., V.D., P.J. and D.Z. thank the {\it International Space Science
  Institute} for facilitating discussions that directly influenced
this work.  
Many Siena College undergraduate students have contributed to this
work, including Corey Snitchler, Trevor Quirk, Erin O'Malley, Amy
McCann, Michael Englert, Debra Johnson, and Alissa Earle.
R.A.K., R.A.F., and M.P.H. acknowledge support of the Undergraduate ALFALFA
Team through NSF AST-1211005, AST-0724918, AST-0725267, and AST-0725380.
The ALFALFA team at Cornell is
   supported by NSF grant AST-1107390 and by the Brinson Foundation. 
This work is based on observations made with the Spitzer Space Telescope, which is operated by 
the Jet Propulsion Laboratory, California Institute of Technology,
under a contract with NASA. 
This research made use of Astropy, a community-developed core Python package for Astronomy (Astropy Collaboration, 2013).
We thank the anonymous referee whose suggestions led to significant
improvements in this paper.

\facilities{Spitzer, IRSA, NED, ADS }

 \bibliography{/Users/rfinn/research/Papers/rfinn}
 \bibliographystyle{aasjournal}

\appendix

\section{GALFIT Simulations}
\label{sim}

The MIPS data have lower resolution and lower signal-to-noise ratios than the optical imaging that
GALFIT is typically used with; at the median redshift of our cluster
sample, one pixel on the MIPS camera corresponds to 1.48~kpc. To test the reliability of the GALFIT models, 
we create 200 model galaxies on each MIPS scan
and run them through our analysis.
The model galaxies consist of single-component \sers \ models with
randomly selected parameters, and the range of
parameters are as follows.  The effective radius is varied uniformly from 1.23 to 20\arcsec, 
the \sers \  index is varied uniformly from 0.5 to 4, and the magnitude
is varied uniformly from 11.5 to 16.  
The model galaxy is created by first selecting 
a region on the MIPS scan that does not have a nearby object within 15
pixels.  We create a cutout of this region, generate a model galaxy
using GALFIT, and then add 
the model and noise to the MIPS cutout.
The PRF may vary across the final MIPS scan; to
incorporate this effect in our simulations, we vary the PRF used to
create the model galaxy by randomly selecting from one of
the five brightest point sources on each image.  
We then run GALFIT on the simulated galaxy to compare the recovered
versus input parameters.  GALFIT requires an initial estimate of the
model parameters; we use the model parameters with 20\% uncertainty 
added to the input parameters, except that we keep the axis ratio and
position angle fixed to reproduce our fitting procedure.  When 
detecting the galaxy, we use the PRF that we use when modeling the
real galaxies (except for Hercules; see below).

\begin{figure*}[h]
\includegraphics[width=.7\textwidth]{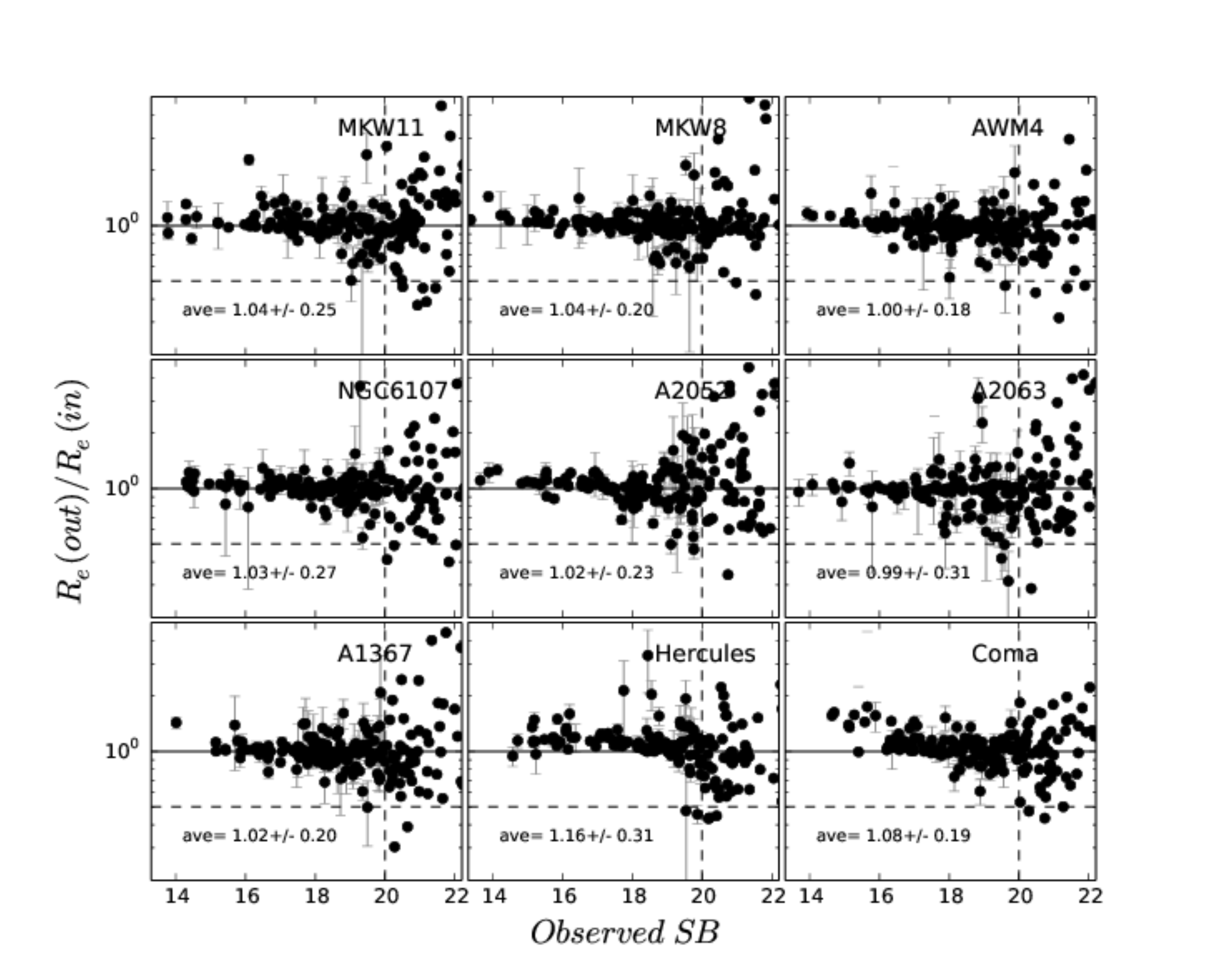}

\caption{\small Ratio of recovered to input $R_e$  of simulated
  galaxies versus observed
  surface brightness.  The solid horizontal light shows a ratio of
  one.  The dashed vertical line shows the surface brightness limit
  applied to each cluster mosaic image, beyond which the fits become
  unacceptably inaccurate.  The average and standard deviation are
  shown in each panel for all galaxies to the left of the dashed
  vertical line.}
\label{GALFITsimsizesb}
\end{figure*}

We show the results of the simulations in Figure
\ref{GALFITsimsizesb}, where we plot the ratio of the recovered to input effective radius versus
measured surface brightness.  
We calculate the observed surface brightness, $\mu$, as
\begin{equation}
\mu = m + 2.5~\log_{10}(\pi R_{e({\rm obs})}^2 ~B/A)
\end{equation}
where m, $R_{e({\rm obs})}$, and $B/A$ are the magnitude, effective radius,
and axis ratio of the best-fit \sers \ model.  (Area of ellipse is $A =
\pi a b$.)  
In each
panel, the scatter in the recovered size increases significantly
beyond a surface brightness of $\sim 20 \ mag/arcsec^2$.
We give the average and standard deviation of the recovered to input $R_e$
for images with $\mu > 20$.  
These are consistent with a ratio of one, indicating that GALFIT is
able to recover the size of the simulated galaxies with $\mu > 20$.

\begin{figure*}[h]
\includegraphics[width=.7\textwidth]{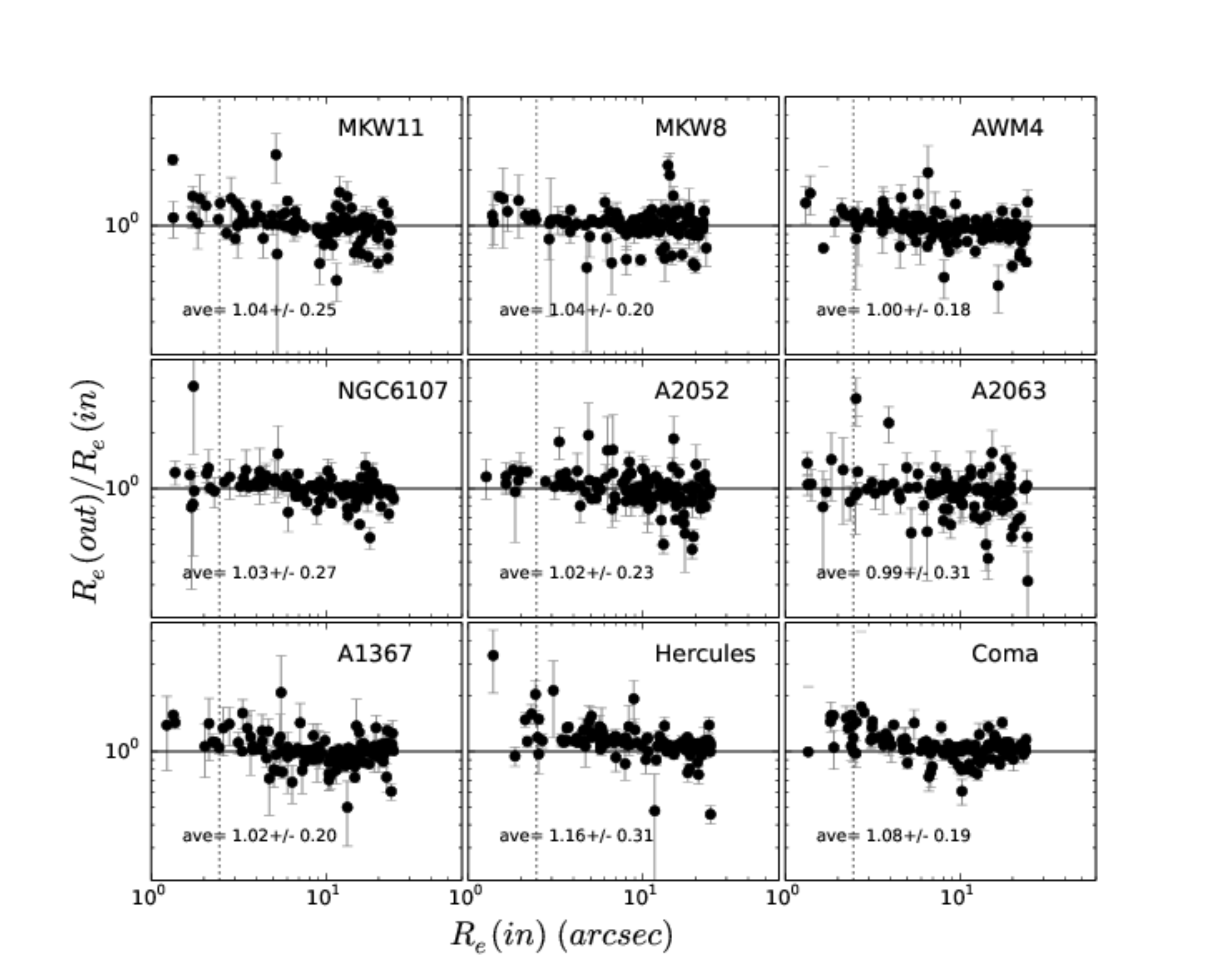}
\caption{\small Recovered versus input galaxy $R_e$ for simulated
  galaxies.  Only galaxies with measured surface brightness $\mu > 20$
  are shown.  The average ratio of recovered to input radius is shown,
and the error is the standard deviation.  For all clusters, with the
exception of Hercules and Coma, the
recovered sizes are consistent with the input sizes.  Hercules, and
Coma to a lesser extent, have recovered sizes that are systematically
larger than the input size.}
\label{GALFITsimRe}
\end{figure*}

In Figure \ref{GALFITsimRe} we show the ratio of the recovered to input effective radius versus
the effective radius of the input model, just for the models that meet
the surface brightness cuts illustrated in Figure \ref{GALFITsimsizesb}.  The vertical
dashed line corresponds to the 2.45\arcsec-pixel scale of the MIPS
scans.  Our convolved models 
are able to recover the input $R_e$ to this limit, although the
recovered values for Hercules are systematically and significantly larger
than the input values.  
When we tried to create a PRF from
a source on the Hercules scan, the recovered radii were $20-30$\%
larger than the input values.  When we use the PRF from $Spitzer \
Science \ Center$, the recovered radii are still systematically high,
but only by 16\%, as shown in Figure \ref{GALFITsimsizesb}.
The remaining offset is likely due to a mismatch in PRF, but we are not
able to further correct for this.  
The simulations indicate that we  will be biased toward measuring larger 24\micron \
$R_e$  in Hercules.

The simulation results lead us to apply a surface brightness cut to
our sample.  In addition to the selection criteria listed in Section
\ref{selection}, we keep
only those galaxies whose measured surface brightness is above $\mu <
20~$mag/sq arcsec.  In physical units, this surface brightness cut
corresponds to $\approx 0.012 ~M_\odot~yr^{-1}~kpc^{-2}$ or $7 \times 10^7
~erg~s^{-1}~ kpc^{-2}$.

\end{document}